\documentclass[twocolumn]{aastex631}
\shorttitle{The Metallicity Gradients of Galaxies at Cosmic Noon in Overdense Environments}
\shortauthors{Li, Wang, Cai et al.}

\usepackage[english]{babel} 
\usepackage[utf8]{inputenc} 
\usepackage[T1]{fontenc}    
\usepackage{ae,aecompl}
\usepackage{pgf,pgfarrows,pgfnodes,pgfautomata,pgfheaps}
\usepackage{graphicx}       
\usepackage{natbib}         
\usepackage{url}            
\usepackage{grffile}        
\usepackage{mathtools}      
\usepackage{multirow}       
\usepackage{xspace}         
\usepackage{booktabs}       

\usepackage{txfonts}
\usepackage{amsmath,amssymb,amsxtra,amsfonts}   

\usepackage{color}         
\definecolor{gold}{rgb}{1,0.80,0}
\definecolor{orange}{rgb}{1,0.5,0}
\definecolor{midgray}{gray}{0.3}
\definecolor{lblue}{rgb}{0,0.2,0.6}
\definecolor{dgreen}{rgb}{0.1,0.6,0.3}
\definecolor{purple}{rgb}{0.5019607843137255,0.0,0.5019607843137255}

\renewcommand\farcs{\mbox{$.\!^{\prime\prime}$}}    
\renewcommand{\arcdeg}{\mbox{$^{\circ}$}\xspace}             

\newcommand{\be}{\begin{equation}}
\newcommand{\ee}{\end{equation}}

\newcommand{\ba}{\begin{align}}
\newcommand{\ea}{\end{align}}

\newcommand{\defeq}{\vcentcolon=}

\newcommand{\Msun}{\ensuremath{M_\odot}\xspace}

\newcommand{\oh}{\ensuremath{12+\log({\rm O/H})}\xspace}

\newcommand{\K}{\ensuremath{\rm K}\xspace}

\newcommand{\Hb}{\textrm{H}\ensuremath{\beta}\xspace}
\newcommand{\Hg}{\textrm{H}\ensuremath{\gamma}\xspace}
\newcommand{\Hd}{\textrm{H}\ensuremath{\delta}\xspace}

\newcommand{\OII}{[\textrm{O}~\textsc{ii}]\xspace}
\newcommand{\OIII}{[\textrm{O}~\textsc{iii}]\xspace}

\newcommand{\NII}{[\textrm{N}~\textsc{ii}]\xspace}

\newcommand{\NeIII}{[\textrm{Ne}~\textsc{iii}]\xspace}
\newcommand{\sersic}{S\'{e}rsic\xspace}
\def\U{\ensuremath{U_{360}}\xspace}     
\def\B{\ensuremath{B_{435}}\xspace}
\def\V{\ensuremath{V_{606}}\xspace}
\def\I{\ensuremath{I_{814}}\xspace}
\def\Y{\ensuremath{Y_{105}}\xspace}
\def\J{\ensuremath{J_{125}}\xspace}

\def\H{\ensuremath{H_{160}}\xspace}

\newcommand{\emc}{\textsc{Emcee}\xspace}

\newcommand{\mg}{\textit{MAMMOTH-Grism}\xspace}

\def\eg{e.g.\xspace}

\renewcommand\({\left(}
\renewcommand\){\right)}

\newcommand{\el}[1]{\ensuremath{\textrm{EL}_{#1}}}

\def\p{{\rm prior}}

\newcommand{\n}{\ensuremath{{\nu}\rm}\xspace}

\received{Feb 5, 2022}
\revised{Mar 15, 2022}
\accepted{Mar 31, 2022}

\graphicspath{{./}{figures/}}
\usepackage{graphicx}
\usepackage{subfloat}

\begin{document}
\email{li-zh21@mails.tsinghua.edu.cn}
\email{wangxin@ipac.caltech.edu}
\email{zcai@mail.tsinghua.edu.cn}
\defcitealias{Forster2018}{F18}
\defcitealias{2022ApJ...926...70W}{W22}
\defcitealias{Wang20}{W20}
\defcitealias{2017ApJ...837...89W}{W17}
\title{First Census of Gas-phase Metallicity Gradients of Star-forming Galaxies in Overdense Environments at Cosmic Noon}

\author[0000-0001-5951-459X]{Zihao Li}
\affiliation{Department of Astronomy, Tsinghua University, Beijing 100084, China}

\author[0000-0002-9373-3865]{Xin Wang}
\affiliation{Infrared Processing and Analysis Center, Caltech, 1200 E. California Boulevard, Pasadena, CA 91125, USA}

\author[0000-0001-8467-6478]{Zheng Cai}
\affiliation{Department of Astronomy, Tsinghua University, Beijing 100084, China}

\author[0000-0002-4314-5686]{Dong Dong Shi}
\affil{Purple Mountain Observatory, Chinese Academy of Sciences, 10 Yuan Hua Road, Nanjing 210023, China}

\author[0000-0003-3310-0131]{Xiaohui Fan}
\affil{Steward Observatory, University of Arizona, 933 North Cherry Avenue, Tucson, AZ 85721, USA}

\author[0000-0003-3728-9912]{Xian~Zhong Zheng}
\affil{Purple Mountain Observatory, Chinese Academy of Sciences, 10 Yuan Hua Road, Nanjing 210023, China}

\author[0000-0001-6919-1237]{Matthew~A.~Malkan}
\affil{Department of Physics and Astronomy, University of California, Los Angeles, CA 90095-1547, USA}

\author[0000-0002-7064-5424]{Harry I. Teplitz}
\affil{Infrared Processing and Analysis Center, Caltech, 1200 E. California Blvd., Pasadena, CA 91125, USA}

\author[0000-0002-6586-4446]{Alaina L. Henry}
\affil{Space Telescope Science Institute, 3700 San Martin Drive, Baltimore, MD, 21218, USA}

\author[0000-0002-8630-6435]{Fuyan Bian}
\affil{European Southern Observatory, Alonso de C\'{o}rdova 3107, Casilla 19001, Vitacura, Santiago 19, Chile}

\author[0000-0001-6482-3020]{James Colbert}
\affil{Infrared Processing and Analysis Center, Caltech, 1200 E. California Blvd., Pasadena, CA 91125, USA}

\begin{abstract}
We report the first spatially resolved measurements of gas-phase metallicity radial gradients in star-forming galaxies in overdense environments at $z\gtrsim2$. The spectroscopic data are acquired by the \mg\ survey, a Hubble Space Telescope (HST) cycle-28 medium program. This program is obtaining 45 orbits of WFC3/IR grism spectroscopy in the density peak regions of three massive galaxy protoclusters (BOSS 1244, BOSS 1542 and BOSS 1441) at $z=2-3$. Our sample in the BOSS 1244 field consists of 20 galaxies with stellar-mass ranging from $10^{9.0}$ to $10^{10.3}$ \Msun\ , star formation rate (SFR) from 10 to 240 \Msun\,yr$^{-1}$, 
and global gas-phase metallicity (\oh) from 8.2 to 8.6. At $1\sigma$ confidence level, 2/20 galaxies in our sample show positive (inverted) gradients --- the relative abundance of oxygen increasing with galactocentric radius, opposite the usual trend.
Furthermore, 1/20 shows negative gradients and 17/20 are consistent with flat gradients. This high fraction of flat/inverted gradients is uncommon in simulations and previous observations conducted in blank fields at similar redshifts.
To understand this, we investigate the correlations among various observed properties of our sample galaxies.
We find an anticorrelation between metallicity gradient and global metallicity of our galaxies residing in extreme overdensities, and a marked deficiency of metallicity in our massive galaxies as compared to their coeval field counterparts.
We conclude that the cold-mode gas accretion plays an active role in shaping the chemical evolution of galaxies in the protocluster environments, diluting their central chemical abundance, and flattening/inverting their metallicity gradients.

\end{abstract}

\section{Introduction} \label{sec:intro}

Metallicity gradients can be a diagnostic of several evolutionary processes in galaxies, including gas inflows and outflows, star formation and evolution \citep{2008MNRAS.385.2181F,10.1093/mnras/stu1288}, and mergers \citep{2010ApJ...710L.156R}. 
Radial metallicity gradients have been under study for decades \citep{1971ApJ...168..327S,1981ARA&A..19...77P,2000A&A...362..921H}. Growing interest in this field has prompted more observations in recent years \citep{Swinbank2012,jones2013,Leethochawalit2016,Forster2018}, together with numerical simulations \citep{2017MNRAS.466.4780M,2019MNRAS.482.2208T,2021MNRAS.506.3024H}. The normal trend of gas-phase metallicity decreases from the inside out, the gradient of which is negative, and we call it inverted if the gradient is positive in the opposite case. Some previous works show negative gradients in local field galaxies \citep{2015MNRAS.448.2030H} as well as at higher redshift \citep[][hereafter W17]{2017ApJ...837...89W}, and positive (inverted) gradients are also found at both local, intermediate ($0.1\lesssim z \lesssim 0.8$) \citep{10.1093/mnras/sty1343}, and higher redshift up to $z\sim3$ \citep{2010Natur.467..811C,2019ApJ...882...94W}. The meaning of the different gradient behaviors is still under debate--more observations, as well as simulations are still needed.

A plausible explanation for inverted gradients is "cold-mode" gas accretion, which has long been recognized to play a crucial role in feeding baryonic gas to galaxies  \citep{2005MNRAS.363....2K,2009Natur.457..451D}. The cold-mode accretion dominates low-mass galaxies (with stellar-mass $M_*\lesssim10^{10.3}\Msun$), where gas from cold dense intergalactic filaments gains gravitational energy and flows into the galaxies at lower temperatures ($T<10^5$K), instead of being shock-heated to the virial temperature of the dark matter (DM) halo ($T\sim10^6\ \mathrm{K}$ for an $M_h\sim10^{12}\Msun$ halo) before cooling and forming stars \citep{2006MNRAS.368....2D}. The latter is conventionally referred to as "hot-mode" accretion, which dominates high-mass galaxies. In the cold-mode accretion scenario, the primordial gas directly flows into the center of galaxies and dilutes the metallicity of the central gas, which disturbs the original distribution of chemical abundance and thus flattens or even inverts the metallicity gradients. The transition between cold and hot mode happens at stellar-mass $M_*\sim10^{10.3}\Msun$ or in terms of halo mass $M_h\sim10^{11.4}\Msun$ \citep{2005MNRAS.363....2K}. However, for galaxies at redshift $z\gtrsim2$, cold-mode accretion could still exist in massive galaxies $M_* \gtrsim 10^{10.3}\Msun$ \citep{2009Natur.457..451D}. 
At high redshift, the cold gas fed by dark matter filaments can penetrate deep into galaxies with halos more massive than the shock-heating scale, where the gas along filaments cools before the pressure develops to support a shock. 

The main observational challenge is that sub-kiloparsec (sub-kpc) resolution (angular resolution $\lesssim0\farcs2$ at $z\sim2$) is required to accurately measure spatial distributions of metallicity in galaxies. Building such a sample for high-z galaxies often suffers from the relatively poor spatial resolution of seeing-limited data, which fails to resolve the inner structures of distant galaxies \citep{10.1093/mnras/sty1343,2020MNRAS.492..821C}. On the other hand, \citet[][hereafter F18]{Forster2018} built a sub-kpc-resolution galaxy sample at $z\sim2$ using SINFONI at the ESO Very Large Telescope (VLT) assisted with adaptive optics. \citet[][hereafter W20]{Wang20} present the first large sub-kpc-resolution sample via grism spectroscopy from the Hubble Space Telescope (HST) in the redshift range of $1.2\lesssim z\lesssim2.3$. \citet{Simons_2021} further extend such analyses to a wider redshift range of $0.6\lesssim z\lesssim2.6$.

In this work, we obtain the first measurements of radial metallicity gradients of galaxies in overdense environments at $z\gtrsim2$, using the data acquired by the MAMMOTH-Grism slitless spectroscopic survey. This survey is a medium program in Hubble Space Telescope Cycle 28 (GO-16276, P.I. Wang), allocated a total of 45 orbits of WFC3/G141 grism spectroscopy and WFC3/F125W pre-imaging in the central field of three of the most massive galaxy protoclusters at cosmic noon. This paper only includes galaxies in the BOSS 1244 protocluster.

This paper is organized as follows. In Section \ref{sec:method}, we describe the grism data reduction, sample selection and metallicity measurements method. We present our finding of the relation between galaxy mass and metallicity gradient in Section \ref{sec:res}. Our conclusions are given in Section \ref{sec:conclusion}. In this paper, we adopt the AB magnitude system, and assume a flat $\Lambda$CDM cosmology with $\Omega_m=0.3,\ \Omega_\Lambda=0.7$, and $H_0=70\ {\rm km\ s^{-1} Mpc}^{-1}$. The metallic lines are indicated in the following manner for brevity, $\OIII\lambda5008\defeq\OIII$, $\OII\lambda\lambda3727,3730\defeq\OII$, $\NeIII~3869\defeq\NeIII$, and $\NII~\lambda6585\defeq\NII$, unless otherwise specified.

\section{Methodology and Measurements}\label{sec:method}

\subsection{Grism Observations and Data Reduction}
The BOSS 1244 protocluster was discovered via the MAMMOTH technique \citep{2016ApJ...833..135C} and spectroscopically confirmed by LBT/MMT IR spectroscopy \citep{Shi2021}. The \textit{Grizli} software package\footnote{\url{https://github.com/gbrammer/grizli/}} is utilized in reducing the paired pre-imaging and grism exposures, following our previous work \citep[][hereafter W22]{2022ApJ...926...70W}. In brief, \textit{Grizli} preprocesses the raw WFC3 exposures by flagging the pixels affected by cosmic rays and persistence, correcting for master and variable sky background caused by the metastable helium glow, and performing WCS alignment to the GAIA DR2 astrometry frame.
After preprocessing, \textit{Grizli} constructs forward-models of the full field-of-view (FoV) grism exposures at the visit level in an iterative manner.
As a result, at the last iteration when the forward-modeling converges, \textit{Grizli} produces a catalog of source redshift and extracted physical properties, e.g., emission-line-fluxes, and spectral indices.
For all sources in this catalog, we also extracted their 1D/2D grism spectra and nebular emission-line 2D postage stamps, with a 60-milliarcsecond plate scale, Nyquist-sampling the WFC3 PSF. 
We use the same data products as in \citetalias{2022ApJ...926...70W}.

\subsection{Sample Selection}
The pre-selection of the sample has been discussed in detail in our previous work \citepalias{2022ApJ...926...70W}. In brief, we fit linear combinations of spectral templates to the optimally extracted 1D grism spectra to infer the grism redshifts. We then compile a sample of 55 spectroscopically confirmed galaxies at $z\sim2.24$, which are likely member galaxies of the BOSS 1244 protocluster \citep{Shi2021}. We also use 1D Gaussian profiles at corresponding wavelength centers to fit their intrinsic nebular emission-lines, and thus obtain emission-line-fluxes (\OIII, \OII, \Hg, \Hb, \Hd). We then select sources with \OIII and \OII emission-lines with both lines having $\mathrm{S/N}\ge3$.
From our diffraction-limited WFC3/G141 spectroscopy of these objects, we obtain their 2D emission-line maps, extracted from their 2D grism spectra after removing their best-fit 2D models of source stellar continua.
We follow the custom technique developed by \citet{Wang20} to deblend the self-contaminating line complex of the \Hb+\OIII$\lambda\lambda$4960,5008 doublets, to produce 2D maps of \OIII$\lambda$5008 and \Hb clean from the orientation-specific contamination of \OIII$\lambda$4960. 

We also considered active galactic nucleus (AGN) ionization, because the strong-line calibrations used to infer metallicity are not valid for the AGN ionization. We rely on the mass-excitation diagram \citep{2014ApJ...788...88J,2015ApJ...801...35C} to exclude AGN candidates in our sample. We identified three galaxies for being likely AGNs (see the right panel of Figure 3 in \citetalias{2022ApJ...926...70W}).

To securely measure metallicity gradients, we apply Voronoi tessellation \citep{2003MNRAS.342..345C} to each galaxy's \OIII map, to bin it into subregions each having $\mathrm{S/N}>2.5$ in \OIII. We further select sources having more than 10 Voronoi bins, in order to have enough points to fit metallicity gradients reliably. Finally, we have 20 galaxies that passed our selection criteria, and 1 galaxy satisfying all criteria but except AGN contamination. We keep the AGN data point in Figure \ref{fig:zg-mass} and Figure \ref{fig:zg-z}, but omit it in other analyses. The integrated metallicity of this AGN is inferred from 2D line-flux maps with the center masked out by a projected $r=1.5$ kpc disk to exclude regions contaminated by the central AGN.

\subsection{Radial metallicity gradients}
\begin{figure*}[ht!]
    \centering
    \plotone{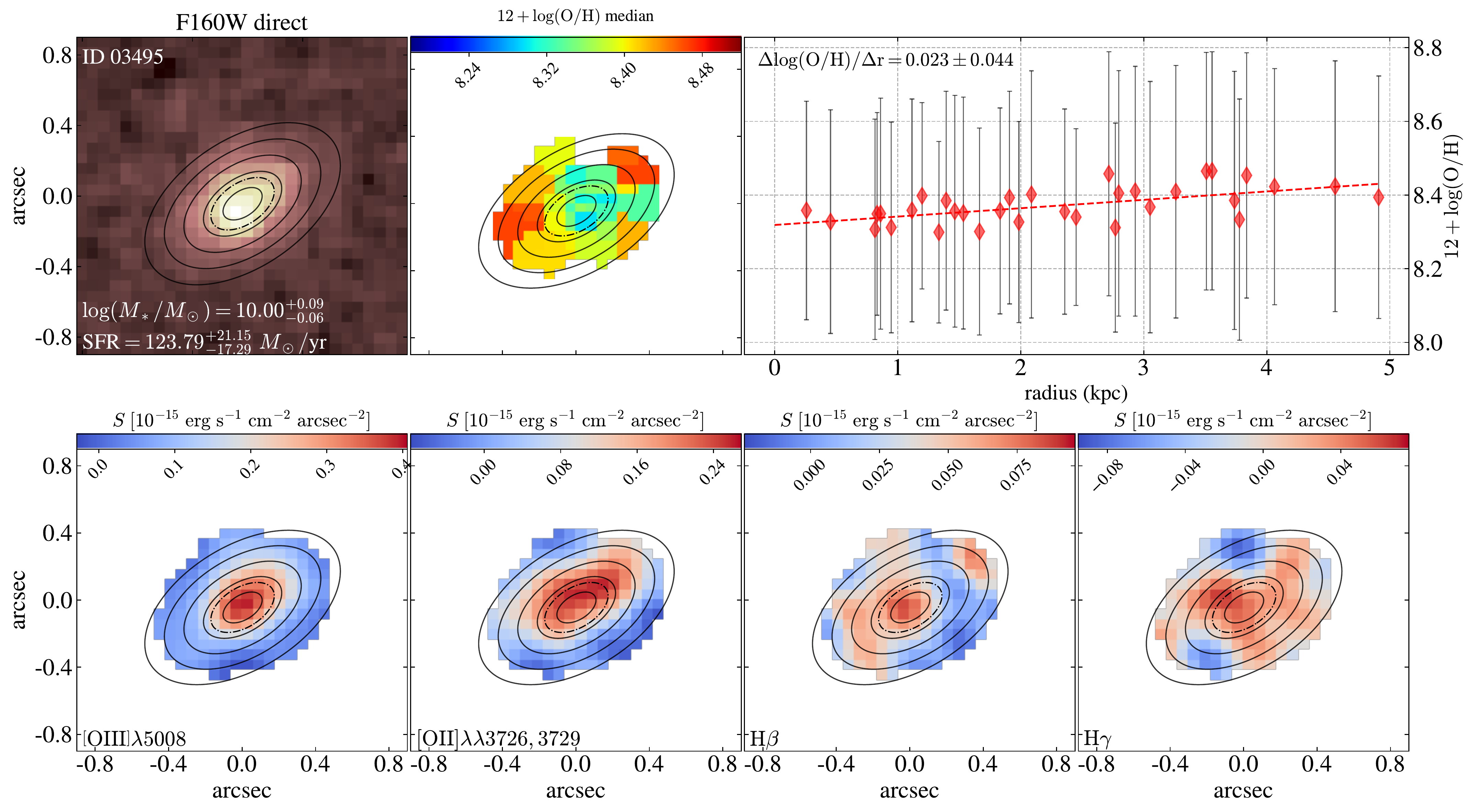}
    \caption{The galaxy ID 03495 as an example of the analysis procedures applied to our entire sample. \textbf{Top}, from left to right: the F160W image of this galaxy, its metallicity map, and radial gradient. The black solid contours mark the deprojected galactocentric radii with a 1 kpc interval. The black dash-dotted ellipses mark the half-light radii fitted by Galfit. \textbf{Bottom}: The smoothed 2D emission-line maps (\OIII,\OII,\Hb,and \Hg) are used to infer the metallicity map. In all maps, north is up and east is to the left.}
    \label{fig:zgrad}
\end{figure*}

Following previous analyses, we assume that all galaxies in our sample are thin rotating disks, supported by recent spatially resolved observations of the ionized gas radial velocity fields at similar redshifts with $M_*\gtrsim10^9\Msun$ \citep[see e.g.][]{2010MNRAS.404.1247J,Forster2018}. The apparent elliptical morphology of galaxies stems from their inclination. Once we fit the inclination angle, we can deproject the images to the source plane and therefore get the intrinsic galactocentric distance. We use the software GALFIT \citep{2002AJ....124..266P} to fit each source morphology. We fit 2D \sersic profiles convolved with the PSF to the direct F160w image. The free parameters in the fitting are the $xy$ coordinates of the galaxy center, the major-axis effective radius $R_e$, the \sersic index $n$, the projected axis ratio $b/a$, and the position angle (PA). We use the $\sigma$ image internally generated by GALFIT in the fitting. For galaxies with irregular shapes (e.g. ID 1435), we use multiple \sersic components in the fitting, and we use the measurements from the brightest component. We get the inclination angle $i$ through the simple relation: $\cos(i)=b/a$. The fitted results for our 20 galaxies are listed in Table \ref{tab:prop}.

We follow procedures in \citetalias{2022ApJ...926...70W} to jointly constrain metallicity (\oh), nebular dust extinction ($A_v$), and de-reddened \Hb flux ($f_{\Hb}$), using our forward-modeling Bayesian inference method. The likelihood function is defined as 
\begin{align}\label{eq:chi2}
    \mathrm{L}\propto\exp\left(-\frac{1}{2}\cdot\sum_i \frac{\(f_{\el{i}} - R_i \cdot 
    f_{\Hb}\)^2}{\(\sigma_{\el{i}}\)^2 + \(f_{\Hb}\)^2\cdot\(\sigma_{R_i}\)^2}\right).
\end{align}
where $f_{\el{i}}$ and $\sigma_{\el{i}}$ represent the de-reddened emission-line (\eg \OII, \Hg, \Hb, \OIII) flux\footnote{ When computing the integrated metallicity, the flux is calculated by fitting multiple Gaussian profiles to the extracted 1D spectra (see Section 3.4 in \citetalias{2022ApJ...926...70W} for details), instead of using 2D emission maps such as when calculating metallicity gradients.} and its uncertainty, corrected using the \citet{2000ApJ...533..682C} dust extinction law with $A_v$ as a free parameter. The emission-line maps are smoothed by a FWHM=0\farcs2 Gaussian kernel, corresponding to the spatial resolution of HST. $R_i$ corresponds to the expected line-flux ratios between EL$_i$ and H$\beta$ (i.e., $R_i$ can be the Balmer decrement of $\Hg/\Hb=0.47$ and the metallicity diagnostics of $\OIII/\Hb$ and $\OII/\Hb$), and $\sigma_{R_i}$ is their intrinsic scatter. The \emc package is employed to perform the MCMC sampling. 

The metallicity measurements are dependent on the specific calibration we choose, because $R_i$ and $\sigma_{R_i}$ are given by strong line calibrations. See \citetalias{2022ApJ...926...70W} for a detailed discussion of the different choices. Here we adopt the empirical strong-line calibration in \citet{bian:18} to convert emission-line-flux ratios to metallicity. 

We apply the metallicity inferred above to each Voronoi bin in galaxies in our sample, so that we can investigate the relation between metallicity and the galactocentric distance. The Voronoi tessellation (as used in \citetalias{Wang20}) is superior to averaging the signal in radial annuli (as is used in \citetalias{Forster2018}), because of azimuthal variations in chemical abundance discovered in nearby spiral galaxies \citep{2017ApJ...846...39H}. We use a simple linear least-squares method to fit the metallicity gradients with the following formula:
\begin{equation}
    \oh=\theta_0+\theta_1 r
\end{equation}
Here $\theta_0$ and $\theta_1$ are the intercept and the gradient of the linear function, respectively, and $r$ is the deprojected galactocentric radius in kiloparsec.

\section{Results} \label{sec:res}
\subsection{Galaxies' Morphology}
We run Galfit as discussed in Section \ref{sec:method} and list the results in Table \ref{tab:prop}. We notice that most of our galaxies have a regular morphology, i.e., they can be well fitted by a single \sersic profile. The exceptions are ID 00313 and ID 01435. They show patterns of off-center clumps. Those may be attributed to star-forming clumps formed by gravitational instabilities in turbulent gas-rich disks \citep{2015Natur.521...54Z}. Considering that the $\chi^2=1.88$ for ID 00313 fitted by Galfit is still acceptable, we retain this result. Because the clump in ID 01435 is more prominent, we therefore add multi-components for this galaxy and use the brightest component to derive the galaxies' morphology properties.

\subsection{Mass dependence of Metallicity Gradients}
\begin{figure*}[t!]
    \centering
    \plotone{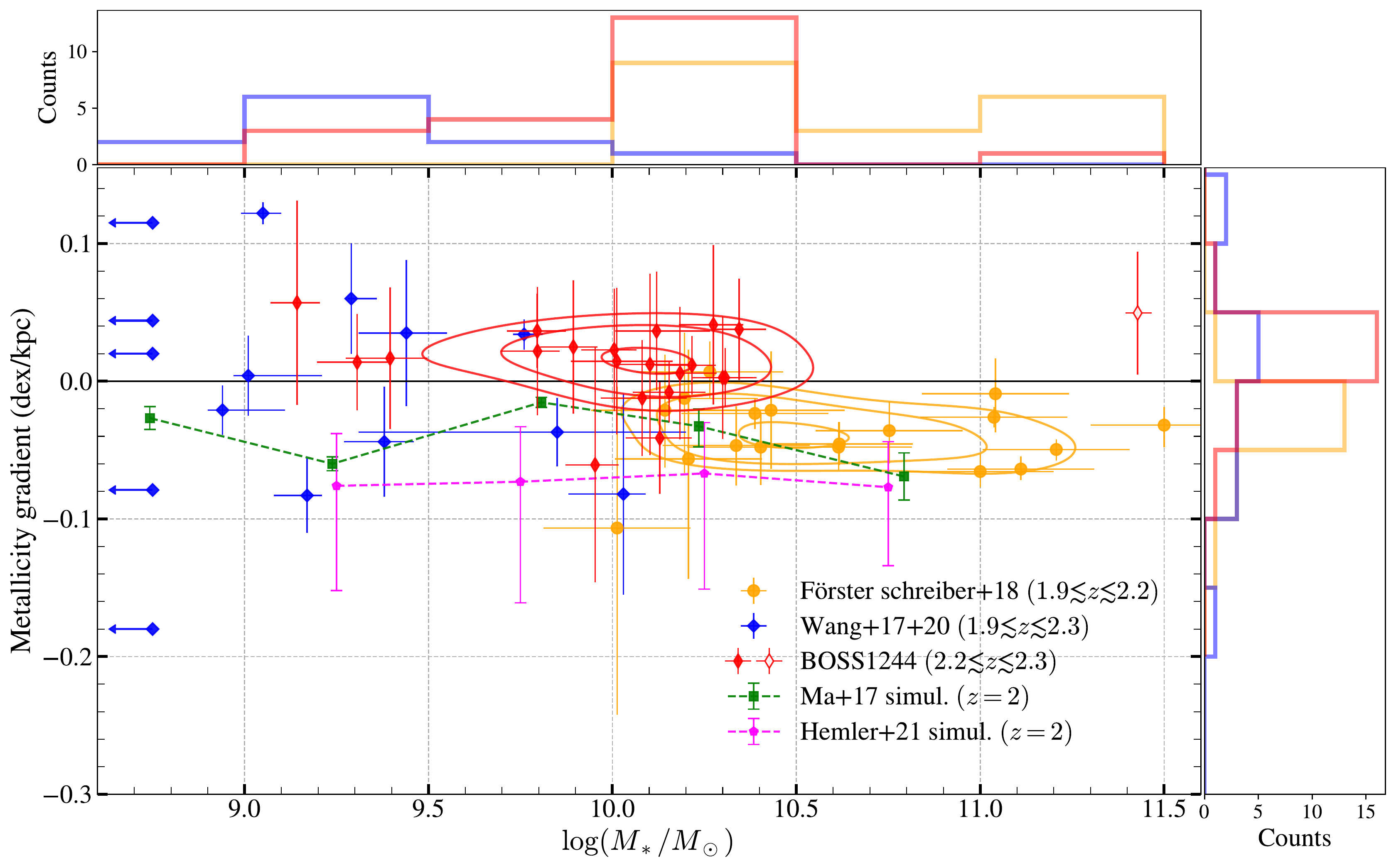}
    \caption{The metallicity gradients as a function of stellar-mass for high-z star-forming galaxies in blank fields and overdense environments. The red diamonds are from our BOSS 1244 field, and the hollow one marks the galaxy categorized as AGN. The orange circles are observed in \citetalias{Forster2018}. The blue diamonds come from the combined sample in \citetalias{2017ApJ...837...89W} and \citetalias{Wang20}, and the ones with left-pointing arrows represent dwarf galaxies ($\log(M_*/M_\odot) <8.5$). These upper limits are omitted from our comparison for being far from our sample mass range. The red and orange contours denote the 50th, 68th, and 95th percentiles of the distribution estimated with Gaussian KDE. The dashed green line is the population stack of five mass bins from FIRE simulations \citep{2017MNRAS.466.4780M}. The magenta dashed line is the population stack of four mass bins from TNG50 simulations \citep{2021MNRAS.506.3024H}. The histograms on the top and right of the main figure show the number distribution of observed stellar-masses and metallicity gradients, respectively. From the histogram, our sample has the largest number of galaxies in the mass range $\log(M_*/M_\odot)=[10,10.5]$ and has a notable number and fraction of inverted metallicity gradients.}
    \label{fig:zg-mass}
\end{figure*}
We list the metallicity gradient measurements in Table \ref{tab:prop}. An example of the fitting together with the galaxy's emission-line maps is shown in Figure \ref{fig:zgrad} and the entire sample is shown in Figure \ref{fig:source-detail} in Appendix \ref{sec:mg-figures}. At the $1\sigma$ confidence level, 2 and 1 galaxies have inverted and negative gradients (i.e., positive or negative and $1\sigma$ away from being flat), while 17 of of the 20 galaxies are consistent with flat gradients. It is surprising that most of our galaxies have flat or marginally inverted gradients, lacking negative gradients, in contrast to expectations from previous cosmological hydrodynamic simulations and observations in blank fields at similar redshifts.

In Figure \ref{fig:zg-mass}, we show the metallicity gradients as a function of the stellar-mass, compared with previous observations in blank fields with sub-kpc resolution and simulations at a similar redshift of $z\sim2$. \citetalias{2017ApJ...837...89W} and \citetalias{Wang20} collect a galaxy sample with a stellar-mass range of $[10^7,10^{10}]\Msun$ using \textit{HST} WFC3 NIR grisms through the Grism Lens-Amplified Survey from Space (GLASS) program; \citetalias{Forster2018} measured galaxies in the mass range of $[10^8,10^{11}]\Msun$ using SINFONI at VLT assisted with adaptive optics. These two samples have robust measurements on radial gradients due to their sub-kpc resolution. There also exist other observations at high redshift \citep{Wuyts_2016,2014MNRAS.443.2695S,2020MNRAS.492..821C}, yet these data are taken under natural seeing.
In simulations, \citet{2017MNRAS.466.4780M} studied a galaxy sample in the mass range $[10^8,10^{11}]\Msun$ from Feedback in Realistic Environments (FIRE) simulations implementing the strong feedback scheme; \citet{2021MNRAS.506.3024H} have a galaxy sample in the mass range $[10^9,10^{11}]\Msun$ from the TNG50 star-forming galaxy population. The predictions of FIRE simulations match well with the negative gradients in \citetalias{Forster2018}. And there are also predominantly negative gradients in the TNG50 simulation. However, our sample shows a large fraction of flat/inverted gradients, where the weighted average metallicity gradient in the BOSS 1244 field is $0.010\pm0.005$; this is hardly seen in \citetalias{Forster2018}, where the weighted average is $-0.041\pm0.004$. The inverted gradients are also found in \citetalias{Wang20}, while our samples have a wider stellar-mass range in $\log(M_*/M_\odot)=[9.5,10.5]$, which strongly suggests that inverted gradients also exist in galaxies with higher masses. 

\begin{figure*}[t]
    \centering
    \plotone{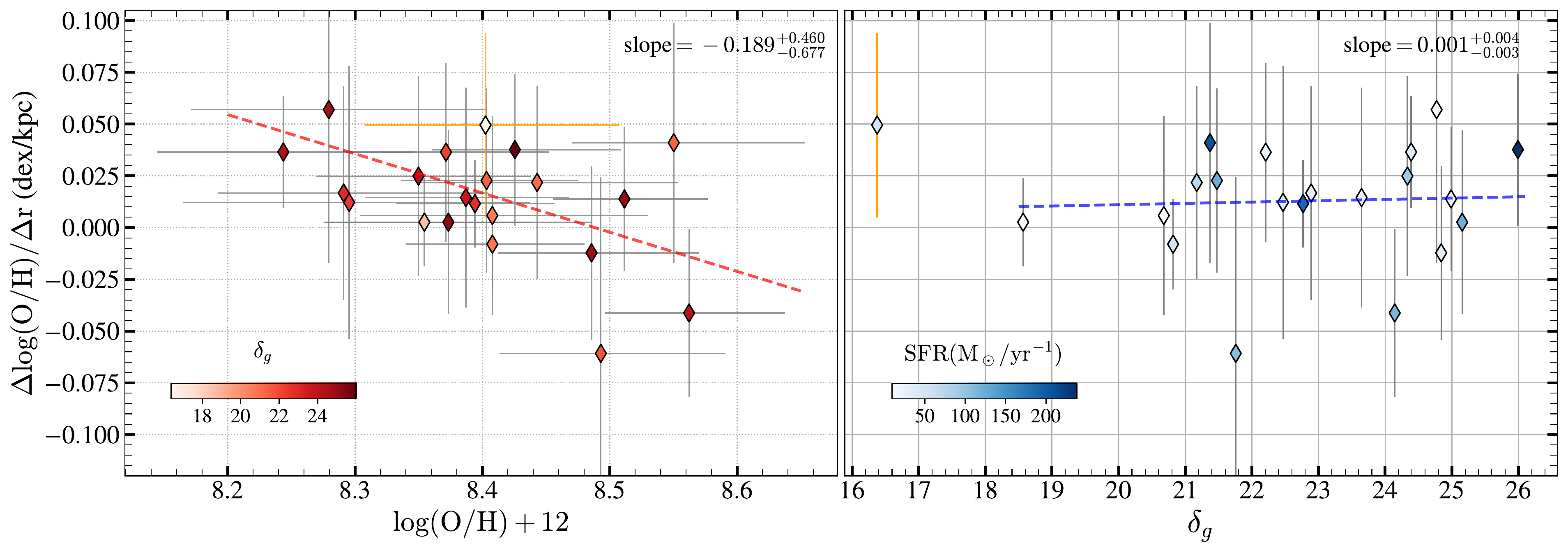}
    \caption{\textbf{Left}: relation between metallicity gradients and global metallicity of our galaxies. The red dashed lines show the linear fitting of the data using the \textsc{linmix} software with $1\sigma$ error of both metallicity and metallicity gradient taken into account. The data points are color-coded by their environment overdensity $\delta_g$ The diamond with an orange error bar marks the galaxy classified as an AGN, which is excluded in the linear fitting in both left and right panels. \textbf{Right}: the metallicity gradients as a function of overdensity $\delta_g$ of our samples. The data points are color-coded in SFR. The blue dashed line is also the linear fitting using \textsc{linmix} taking into account $1\sigma$ errors of the metallicity gradients. The data points are color-coded in SFR. }
    \label{fig:zg-z}
\end{figure*}
\subsection{Environmental effects}

\begin{figure}[h!]
    \centering
    \plotone{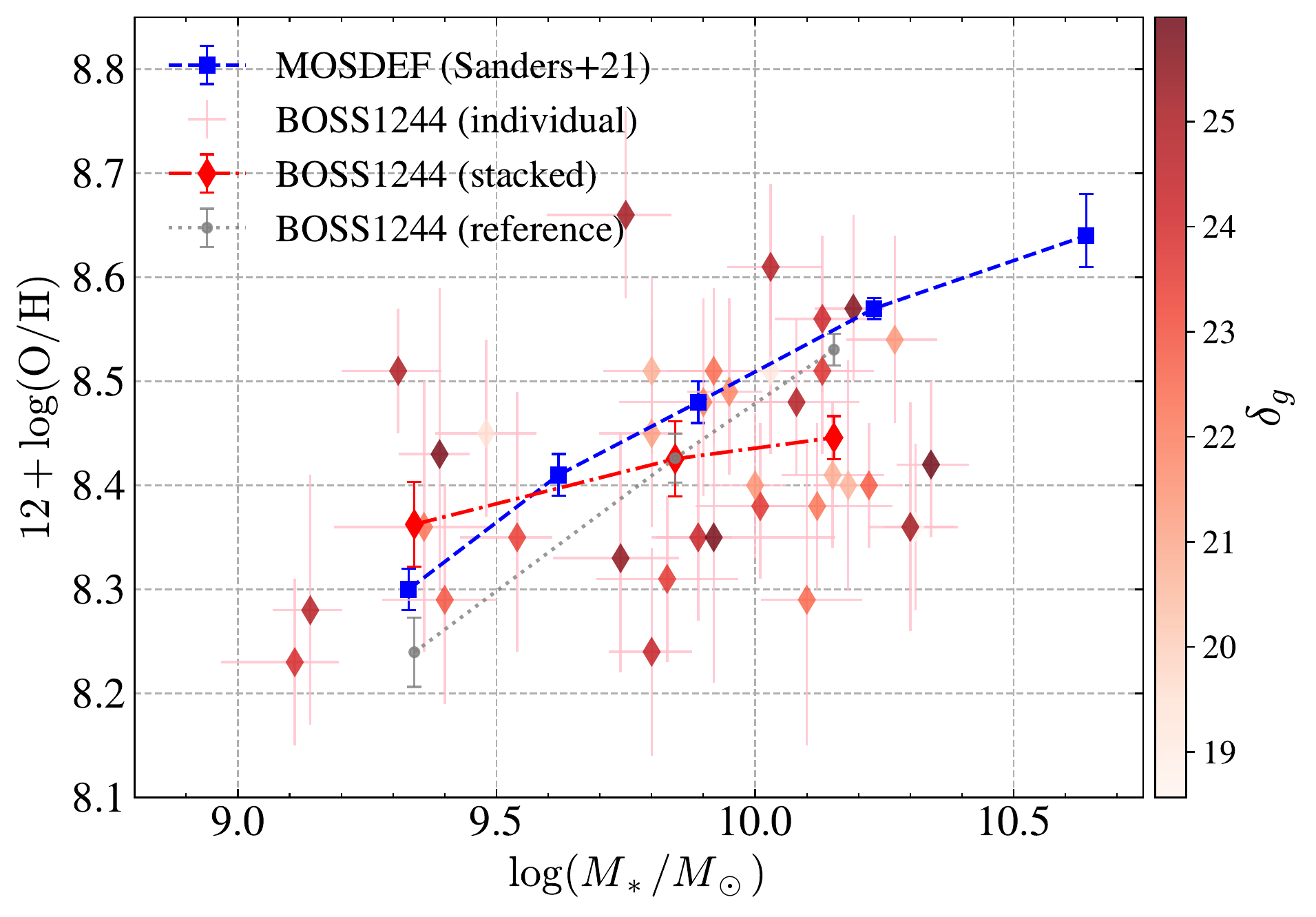}
    \caption{The mass-metallicity relation (MZR) for our galaxy sample  compared with the MZR measured in the field by the MOSDEF survey at $z\sim2.3$ \citep{2021ApJ...914...19S}. The diamonds with pink bars represent individual galaxies, color-coded in overdensity $\delta_g$. The red and blue lines correspond to the stacks of galaxies in our BOSS 1244 protocluster and the MOSDEF fields. The gray dotted line shows the reference MZR derived using the \citet{2021ApJ...914...19S} FMR assuming the SFR and $M_*$ of our sample galaxies.
    We see that our galaxies residing in overdense environments show significant metallicity deficiency, compared with the their coeval ﬁeld counterparts, particularly in the high-mass regime. Here we 
    not only show the 20 galaxies in this work, but also the other 16 galaxies in \citetalias{2022ApJ...926...70W}, to illustrate the relation in our protocluster environment with stronger statistical significance.}
    \label{fig:z-mass}
\end{figure}
We plot metallicity gradients as a function of the integrated galaxy metallicity on the left panel of Figure \ref{fig:zg-z}. We see evidence for an anticorrelation between metallicity and metallicity gradient (hereafter Zgrad-Z relation for simplicity), with Pearson coefficient $r=-0.49$ and p-value $p=0.03$. This low p-value indicates that the metallicity and metallicity gradients are correlated at the $95\%$ confidence level. We find the metal-rich galaxies tend to show negative metallicity gradients, while metal-poor galaxies are more likely to have inverted gradients. It is also consistent with the galaxy mass-dependence of the metallicity gradient, in the sense that more massive galaxies tend to be more metal-rich \citep{Salim_2015} and metallicity gradient is anticorrelated with mass \citep{2017MNRAS.466.4780M,Wang20}. 

When primordial gas flows into a galaxy, it dilutes the central metal abundance, raises the metallicity gradient, and, at the same time, also dilutes the integrated metallicity of the galaxy. This can explain why we tend to see stronger inverted metallicity gradients for metal-poor galaxies. In Figure \ref{fig:z-mass}, we show the mass-metallicity relation (MZR) of our 20 galaxies residing in extreme overdensities and that from the MOSDEF survey targeting blank fields, at similar redshifts ($z\sim2.3$) \citep{2021ApJ...914...19S}. The comparison of the MZRs in our BOSS 1244 high-density field and the MOSDEF blank fields has already been discussed in \citetalias{2022ApJ...926...70W}, and here we only focus on galaxies with metallicity gradient measurements. 
We caution that these surveys have different line-flux limits, which results in different source selections that can affect the behavior of the MZR \citep[see a recent comprehensive analysis by][]{Henry:2021ju}.
To account for this selection effect, we employ the fundamental metallicity relation (FMR) prescribed by \citet{2021ApJ...914...19S} to derive the field reference values of metallicities for our galaxies. 

We find a marked deficiency of metallicity for our galaxies, in particular with $M_*\sim10^{10}\Msun$, as compared to their field reference values.
This metallicity deficiency is possibly caused by the combined effects of cold-mode gas accretion from cosmic filaments \citep{Keres2009} and metal-enriched gas removal by ejective feedback \citep{Muratov:2015er}.
\citet{2017MNRAS.466.4692K} found that massive galaxies, with their deep gravitational potentials, should accrete cold IGM gas from the intrafilament medium. We suggest that the gravitational potential of the cluster-scale dark matter halo in overdense environments would also draw primordial gas from the IGM through cosmic filaments. This cold-mode accretion boosted by the extremely massive protocluster dominates our high mass galaxies, diluting the metallicity, making them more metal-poor than their coeval field counterparts. The star-forming feedback blows metal-rich winds outward and flattens/inverts the metallicity gradients \citep{2019ApJ...882...94W}. For energy-driven winds, the mass loading factor is $\eta\propto v_{vir}^{-2}\propto M_h^{-1/3}$, and $\eta\propto v_{vir}^{-1}\propto M_h^{-2/3}$ for momentum-driven winds. Thus, the fraction of metal-rich gas that is blown outward by feedback decreases with halo mass, so the outflows are suppressed in high-mass galaxies due to their massive DM halos, as well as the cluster-scale DM halo. In this case, the cold-mode accretion enhanced by the protocluster environment should dominate for most of the galaxies, while star-formation feedback that is suppressed by DM halos would play a minor role.

On the right panel of Figure \ref{fig:zg-z}, we plot the relation between metallicity gradient and galaxy overdensity $\delta_g$, where $\delta_g$ is defined as $\delta_g=\frac{\Sigma_{\rm group}}{\Sigma_{\rm group}}-1$, $\Sigma_{\rm group}$ is the H$\alpha$ emitter (HAE) number per arcmin$^2$ within the overdensity, and $\Sigma_{\rm field}$ is the surface density of HAEs in random fields. We estimate the $\delta_g$ field by fitting 2D kernel density estimation (KDE) to the HAE map with a $200^{''}(\sim5.3\ c\rm{Mpc})$ Gaussian kernel, normalized by the maximum $\delta_g$ of the group calculated in \citet{Shi2021}. We also use \textsc{Linmix} to fit a linear relation. We find no significant correlation between metallicity gradient and $\delta_g$, with a Pearson coefficient $r=0.17$ and $p$-value $p=0.46$. The protocluster may be too dense to reach a certain threshold (all sample galaxies reside in a similarly overdense environment, far denser than blank fields), at which there is enough gas to replenish all the member galaxies; thus, the metallicity mainly depends on each galaxy's individual properties. Thus, the trend that metallicity gradients grow as overdensity increases is not so obvious. We expect a more significant correlation as the galaxy sample continues to accumulate in the \mg\ survey. 
\begin{longrotatetable}
\begin{deluxetable*}{cccccccccccccccccc}   \tablecolumns{18}
\tabletypesize{\scriptsize}
\tablewidth{0pt}
\tablecaption{Derived physical and morphological properties of the galaxies\label{tab:prop}}
\tablehead{
    \colhead{ID} & 
    \colhead{R.A.} & 
    \colhead{Decl.} & 
    \colhead{$z_\mathrm{grism}$} & 
    \colhead{$\delta_g$} &
    \multicolumn{5}{c}{Photometry (ABmag)} &
    \multicolumn{4}{c}{Morphology Properties} &
    \multicolumn{4}{c}{Derived Physical Properties} \\
    & (deg.) & (deg.) & & & \multicolumn{5}{c}{\hrulefill}  &  \multicolumn{4}{c}{\hrulefill} & 
    \multicolumn{4}{c}{\hrulefill} \\
     & & & & & \colhead{F125W} & \colhead{F160W} & \colhead{U} & \colhead{$z$} & \colhead{$K_s$} & \colhead{$R_e$} & \colhead{$b/a$} & \colhead{PA} &$\chi^2$ & \colhead{$\log(M_*/\Msun)$} & \colhead{SFR (\Msun $yr^{-1}$) \tablenotemark{a}} & \colhead{\oh} & \colhead{$\nabla Z$ [$\mathrm{dex\ kpc^{-1}}$]}
}
\startdata
00262\tablenotemark{b} & 190.938355 & 35.849177 & 2.22 & 16.38 & 22.59 & 21.83 & 24.71 & 23.04 & 20.50 & 1.95 & 0.75 & 16.95 & 1.33 & $11.43^{+0.03}_{-0.04}$ & $51.01^{+9.03}_{-8.51}$ & $8.40^{+0.09}_{-0.11}$ & $0.050\pm0.045$ \\
00313 & 190.902894 & 35.851659 & 2.23 & 18.57 & 24.45 & 23.68 & 25.50 & 24.21 & 22.61 & 2.63 & 0.29 & 17.07 & 1.88 & $10.31^{+0.06}_{-0.08}$ & $10.64^{+1.27}_{-1.52}$ & $8.35^{+0.08}_{-0.08}$ & $0.003\pm0.021$ \\
00613 & 190.921119 & 35.865813 & 2.24 & 20.68 & 24.30 & 23.78 & 25.38 & 24.37 & 22.82 & 2.24 & 0.44 & 40.15 & 1.20 & $10.18^{+0.07}_{-0.10}$ & $9.32^{+1.01}_{-1.03}$ & $8.41^{+0.10}_{-0.12}$ & $0.006\pm0.048$ \\
00769 & 190.909740 & 35.871348 & 2.24 & 22.21 & 23.43 & 23.05 & 24.43 & 23.35 & 22.25 & 2.36 & 0.71 & -14.18 & 1.39 & $10.12^{+0.11}_{-0.15}$ & $23.06^{+4.72}_{-4.33}$ & $8.37^{+0.08}_{-0.08}$ & $0.036\pm0.043$ \\
00996 & 190.873758 & 35.880726 & 2.32 & 22.47 & 23.96 & 23.41 & 25.86 & 24.12 & 23.04 & 1.14 & 0.50 & -63.54 & 1.22 & $10.10^{+0.09}_{-0.11}$ & $38.61^{+10.12}_{-6.78}$ & $8.30^{+0.13}_{-0.14}$ & $0.012\pm0.066$ \\
01335 & 190.871757 & 35.895697 & 2.24 & 24.14 & 23.04 & 22.72 & 24.41 & 22.81 & 21.73 & 4.11 & 0.43 & 86.05 & 1.34 & $10.13^{+0.09}_{-0.09}$ & $114.05^{+22.49}_{-29.10}$ & $8.56^{+0.07}_{-0.08}$ & $-0.041\pm0.040$ \\
01394 & 190.877806 & 35.898194 & 2.21 & 24.99 & 23.81 & 23.61 & 24.84 & 23.67 & 22.71 & 0.99 & 0.40 & 24.13 & 1.31 & $9.31^{+0.11}_{-0.08}$ & $24.75^{+4.79}_{-4.45}$ & $8.51^{+0.06}_{-0.07}$ & $0.014\pm0.035$ \\
01435 & 190.869813 & 35.900036 & 2.21 & 24.33 & 23.46 & 23.13 & 25.35 & 23.24 & 22.27 & 1.02 & 0.97 & -5.47 & 1.25 & $9.89^{+0.09}_{-0.07}$ & $93.26^{+15.22}_{-12.83}$ & $8.35^{+0.08}_{-0.09}$ & $0.025\pm0.048$ \\
01464 & 190.873403 & 35.901270 & 2.21 & 24.84 & 23.81 & 23.40 & 25.83 & 24.01 & 22.49 & 1.31 & 0.43 & 74.13 & 1.34 & $10.08^{+0.11}_{-0.12}$ & $33.40^{+6.59}_{-5.93}$ & $8.49^{+0.07}_{-0.08}$ & $-0.012\pm0.042$ \\
01467 & 190.869254 & 35.901455 & 2.21 & 24.39 & 24.00 & 23.51 & 25.52 & 23.88 & 23.43 & 1.68 & 0.40 & -71.41 & 1.40 & $9.80^{+0.08}_{-0.08}$ & $34.16^{+9.66}_{-6.89}$ & $8.24^{+0.10}_{-0.10}$ & $0.037\pm0.027$ \\
01890 & 190.865939 & 35.913638 & 2.25 & 24.77 & 25.08 & 24.36 & 26.36 & 24.43 & 24.36 & 1.93 & 0.46 & 17.43 & 1.17 & $9.14^{+0.07}_{-0.06}$ & $17.05^{+2.71}_{-2.40}$ & $8.28^{+0.11}_{-0.12}$ & $0.057\pm0.074$ \\
01998 & 190.868202 & 35.916375 & 2.21 & 25.16 & 22.83 & 22.49 & 23.94 & 22.26 & 21.25 & 1.89 & 0.95 & -37.99 & 1.55 & $10.30^{+0.08}_{-0.09}$ & $121.59^{+37.28}_{-26.68}$ & $8.37^{+0.10}_{-0.11}$ & $0.003\pm0.044$ \\
02327 & 190.927963 & 35.926164 & 2.24 & 23.65 & 23.99 & 23.52 & 25.93 & 24.25 & 22.82 & 1.08 & 0.65 & 57.80 & 1.22 & $10.01^{+0.13}_{-0.15}$ & $24.76^{+4.96}_{-4.73}$ & $8.39^{+0.08}_{-0.08}$ & $0.014\pm0.053$ \\
02330 & 190.905581 & 35.926178 & 2.29 & 25.99 & 22.78 & 22.39 & 25.42 & 22.89 & 21.27 & 1.31 & 0.97 & 23.60 & 1.29 & $10.34^{+0.07}_{-0.07}$ & $238.14^{+24.23}_{-24.36}$ & $8.43^{+0.07}_{-0.08}$ & $0.038\pm0.037$ \\
02992 & 190.843993 & 35.949979 & 2.35 & 21.76 & 23.84 & 23.29 & 26.40 & 23.67 & 22.97 & 1.37 & 0.72 & -86.93 & 1.30 & $9.95^{+0.08}_{-0.06}$ & $109.41^{+17.00}_{-15.95}$ & $8.49^{+0.08}_{-0.10}$ & $-0.061\pm0.085$ \\
03061 & 190.838668 & 35.953087 & 2.23 & 20.82 & 23.07 & 22.73 & 24.39 & 22.90 & 21.82 & 2.79 & 0.36 & -36.25 & 1.25 & $10.15^{+0.10}_{-0.10}$ & $49.78^{+9.25}_{-8.66}$ & $8.41^{+0.07}_{-0.07}$ & $-0.008\pm0.022$ \\
03155 & 190.842877 & 35.957764 & 2.32 & 21.37 & 23.48 & 23.02 & 26.32 & 23.60 & 22.00 & 2.44 & 0.80 & 58.41 & 1.30 & $10.27^{+0.09}_{-0.08}$ & $207.88^{+30.67}_{-28.99}$ & $8.55^{+0.08}_{-0.10}$ & $0.041\pm0.058$ \\
03276 & 190.857451 & 35.964185 & 2.21 & 22.77 & 22.48 & 22.18 & 24.04 & 22.20 & 21.43 & 3.13 & 0.54 & -9.10 & 1.51 & $10.22^{+0.10}_{-0.07}$ & $189.85^{+30.04}_{-25.10}$ & $8.39^{+0.06}_{-0.06}$ & $0.012\pm0.021$ \\
03331 & 190.862922 & 35.967686 & 2.22 & 22.89 & 24.04 & 23.71 & 25.20 & 23.71 & 22.76 & 1.78 & 0.34 & -6.72 & 1.23 & $9.40^{+0.12}_{-0.10}$ & $28.11^{+4.40}_{-5.40}$ & $8.29^{+0.10}_{-0.11}$ & $0.017\pm0.052$ \\
03495 & 190.853920 & 35.977367 & 2.23 & 21.48 & 23.72 & 23.22 & 26.49 & 23.70 & 22.86 & 1.62 & 0.59 & -53.70 & 1.22 & $10.00^{+0.09}_{-0.06}$ & $123.79^{+21.15}_{-17.29}$ & $8.40^{+0.07}_{-0.07}$ & $0.023\pm0.044$ \\
03516 & 190.851775 & 35.979114 & 2.23 & 21.17 & 23.79 & 23.46 & 25.86 & 23.63 & 22.71 & 2.50 & 0.45 & -27.16 & 1.20 & $9.80^{+0.10}_{-0.06}$ & $75.17^{+13.24}_{-9.98}$ & $8.44^{+0.10}_{-0.11}$ & $0.022\pm0.047$ \\
\enddata
\tablecomments{
\tablenotetext{a}{The SFR is derived through SED fitting using J-(HST F125W), H-(HST F160W), U-(LBT/LBC Uspec), $z$-(LBT/LBC $z$-SLOAN) and $K_s$-band (CFHT $K_s$) photometry, assuming a constant star formation history, but the SFR in our previous work \citepalias{2022ApJ...926...70W} is derived through dust-corrected \Hb flux.}
\tablenotetext{b}{This object is classified as an AGN. Its integrated metallicity is inferred from 2D line-flux maps with the center masked out by a deprojected $r=1.5$ kpc disk, to discard regions contaminated by the central AGN, whereas the line-fluxes for objects without AGNs are estimated from fitting 1D Gaussian profiles in 1D spectra.}
}
\end{deluxetable*}
\end{longrotatetable}

\section{Conclusion}\label{sec:conclusion}
We have presented the first sample of gas-phase metallicity radial gradients measured in overdense environments at $z>2$ using grism slitless spectroscopy. The data presented in this work were acquired by \mg\ in HST cycle-28 medium program. The BOSS 1244 protocluster field is among three of the most massive galaxy protoclusters at $z \sim 2.2-2.3$ identified using the MAMMOTH technique, in which we selected a sample of 20 protocluster member galaxies with $M_*/\Msun\in[10^9,10^{10.3}]$, SFR $\in[10,240]\Msun\rm{yr}^{-1}$ and $\oh\in[8.2,8.6]$ to measure metallicity gradients. We find an unprecedentedly large fraction of flat/inverted gradients in this protocluster environment, compared with that of MOSDEF field galaxies. At the $1\sigma$ confidence level, we find 2/20 of galaxies showing inverted gradients and 17/20 with no significant gradients. This differs from the usual trend of negative gradients. We find an anticorrelation between metallicity gradients and integrated metallicity. We conclude that these notable flat/inverted gradients are likely caused by strong cold-mode accretion in protoclusters. The overdense environments boost cold-mode accretion by dragging more primordial gas through cosmic filaments and injecting it into the galaxies’ centers through the strong gravitation of the massive cluster-scale DM halo. The primordial gas directly dilutes the integrated metallicity and also flattens/inverts the metallicity gradients. This scenario is consistent with the anticorrelation between metallicity gradients and integrated metallicity, where we expect the galaxy to be more metal poor when it shows a higher metallicity gradient in response to gas diluting its center.

While the hydrodynamical simulations so far match the negative metallicity gradients well for field galaxies, they cannot reproduce our findings in overdense environments. This also suggests that future hydrodynamical simulations should take such environmental effects into consideration to investigate the galaxy mass assembly processes in protoclusters.

The ongoing \mg\ program includes the other two protocluster fields, BOSS 1542 and BOSS 1441 \citep{2017ApJ...839..131C,Shi2021}. We expect our statistics will be improved once all the data are acquired by HST and analyzed, when the sample size will be three times as large as now. This \mg\ full sample of galaxies at cosmic noon in extremely overdense environments provides a unique opportunity for us to further understand environmental effects on galaxy formation.

\section{Acknowledgements}

We would thank the anonymous referee for helpful comments that ensure the quality of this paper. Z.L. and Z.C. are supported by the National Key RD Program of China
(grant No. 2018YFA0404503), the National Science Foundation of China (grant No. 12073014), and the science research grants from the China Manned Space Project with NO. CMS-CSST-2021-A05. This work is supported by NASA through HST grant HST-GO-16276.

\software{\emc \citep{2013PASP..125..306F}, Grizli \citep{2021zndo...5012699B}, GALFIT \citep{2002AJ....124..266P}, VorBin \citep{2003MNRAS.342..345C}, \textsc{Linmix} \citep{2007ApJ...665.1489K}.}
\appendix

\section{Systematic error estimation using mocks}

To test the robustness of our method and estimate the systematic error, we use mocks to compare the reconstructed metallicity gradients with the input values. To build a simplified model of emission-line maps, we consider a face-on galaxy with \Hb having a S\'{e}rsic profile with index $n=0.5$ and $R_e=1.5$kpc. We then assign the metallicity to the disk with given gradients, and obtain other line maps (\OIII, \OII, and \Hg) from the \Hb maps considering by the metallicity calibrations in \citet{bian:18}. We add Gaussian random noise to the line maps so that they have the desired $\mathrm{S/N}$. We then apply the pipeline discussed in our paper to the mock emission-line maps to calculate the metallicity gradients. In the mocks, we choose five metallicity gradients: $[0,0.02,0.04,-0.02,-0.04]\ \mathrm{dex\ kpc^{-1}}$, and $(\mathrm{S/N})_{\rm H_\beta}=3$, comparable to the median $(\mathrm{S/N})_{\rm H_\beta}=3.4$ of our sample galaxies.

\begin{figure*}[h]
    \plotone{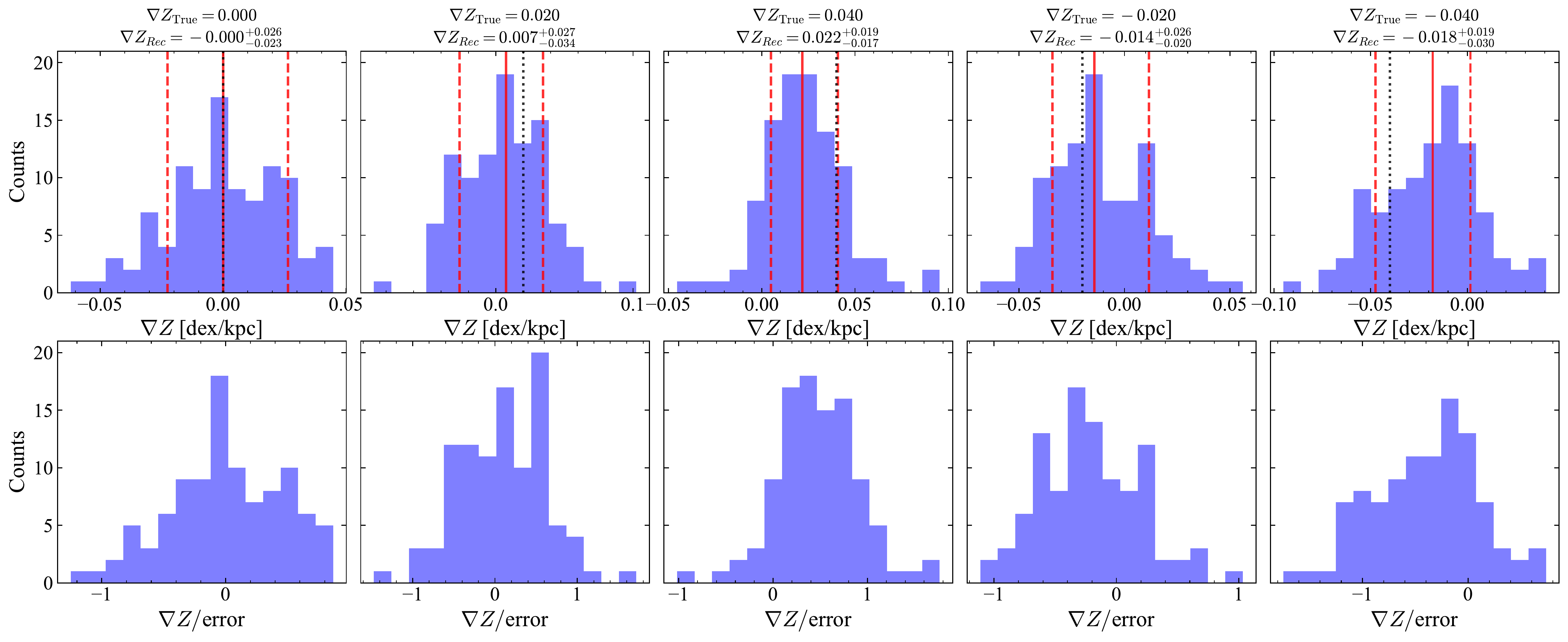}
    \caption{The histogram of the fitted metallicity gradients (top), and the metallicity gradients divided by their uncertainties (bottom). From left to right: the results for five different mocks with true gradients $[0,0.02,0.04,-0.02,-0.04]\ \mathrm{dex\ kpc^{-1}}$. The black dotted line denotes the true gradient value. The solid red line represents the median value in the 100 reconstructions, and the red dashed lines show the $1\sigma$ interval. \label{fig:mock}}
\end{figure*}

We run each case 100 times and estimated the systematic error to be $\sim0.02\ \mathrm{dex\ kpc^{-1}}$. From the first panel of Figure \ref{fig:mock}, we find that the reconstructed metallicity gradients generally agree with the true values within $1\sigma$ in all five cases. We also note that the reconstructed gradients tend to be flat: the reconstructed gradients are less steep than the original input values for both positive and negative cases. From the second panel, we find that most of the measurements are below the $1\sigma$ significance level, similar to the situation of our measurements in the BOSS 1244 field. 

Interestingly, in the two mocks with positive gradients, we find that the ratio of positive reconstructed gradients are $[59\%,87\%]$ for $\nabla Z_{\rm True}=[0.02,0.04]\ \mathrm{dex\ kpc^{-1}}$ respectively, while in our galaxy sample, the ratio of positive gradients is $16/20=80\%$. This hints that our sample galaxies tend to have intrinsic positive gradients, despite the relatively large uncertainty in metallicity calculation and gradient fitting.

\section{Detailed figures for measuring radial gradients} \label{sec:mg-figures}

\begin{figure}[htbp]
    \centering
    
    \gridline{
            \fig{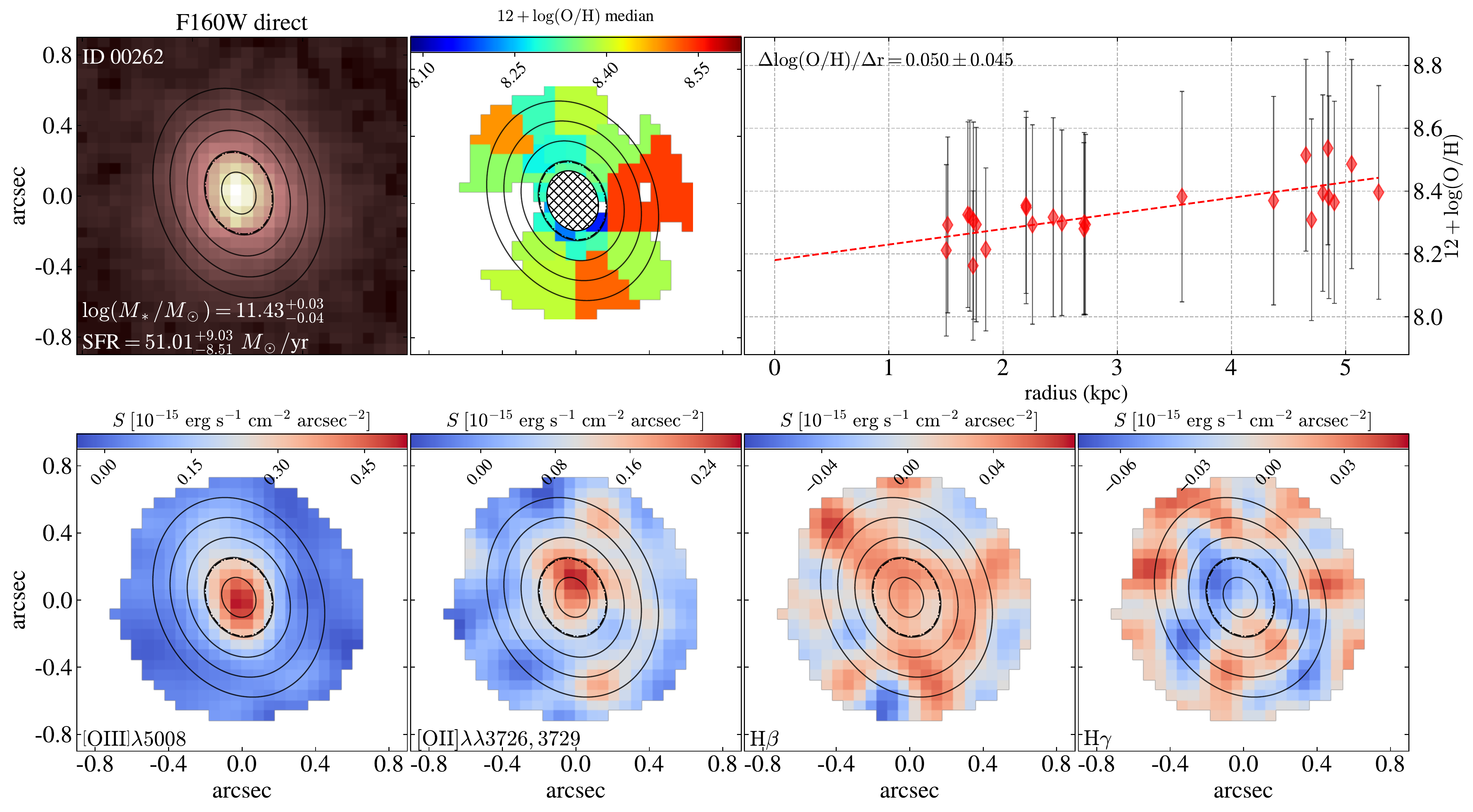}{0.495\textwidth}{ID 00262 (AGN)}
            \fig{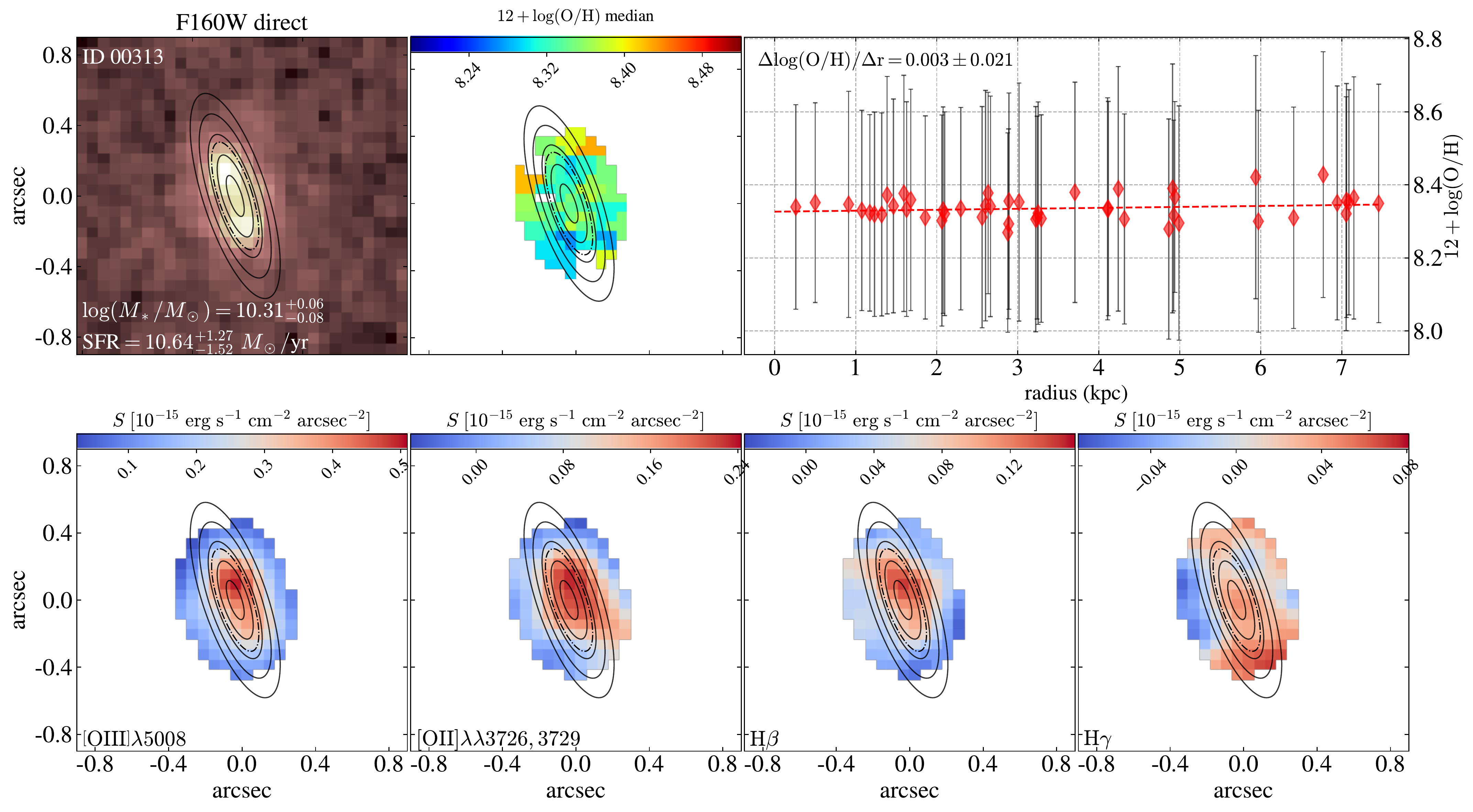}{0.495\textwidth}{ID 00313}
            }
    
    \gridline{
            \fig{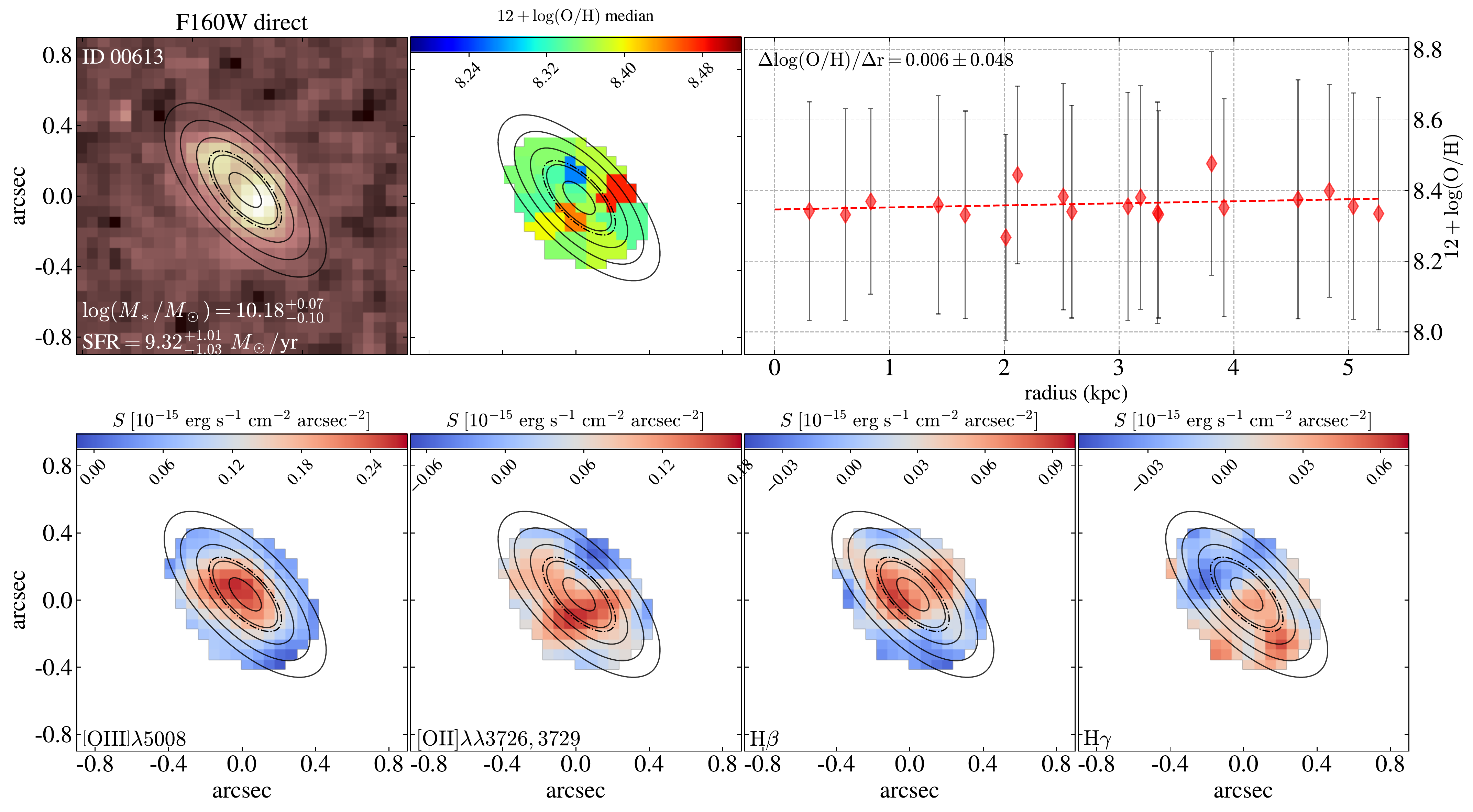}{0.495\textwidth}{ID 00613}
            \fig{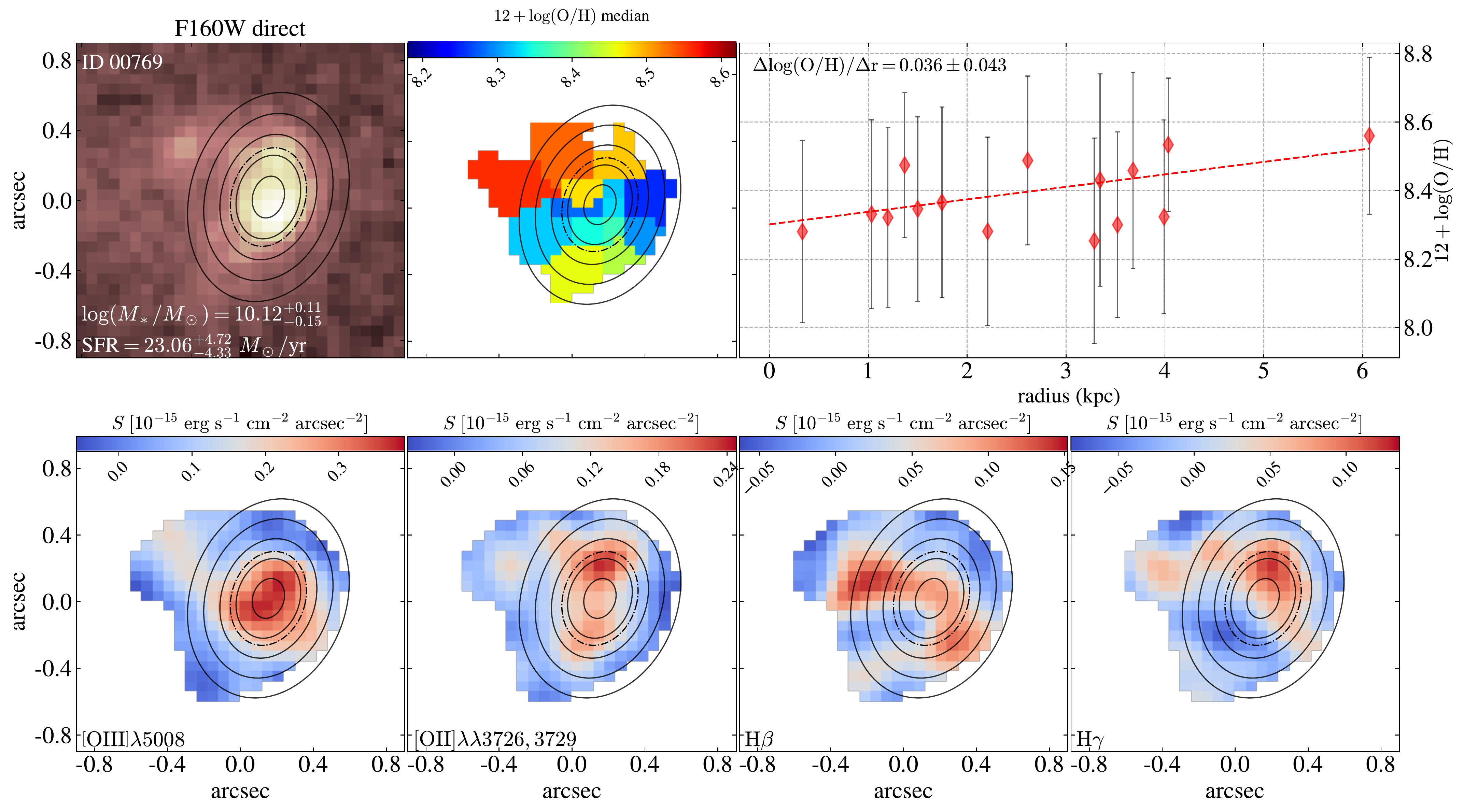}{0.495\textwidth}{ID 00769}
            }

\caption{Metallicity radial gradients measurement in our sample. The same as Figure \ref{fig:zgrad}}
\label{fig:source-detail}
\end{figure}
\addtocounter{figure}{-1}

\begin{figure}[htbp]
    \centering
    \gridline{
        \fig{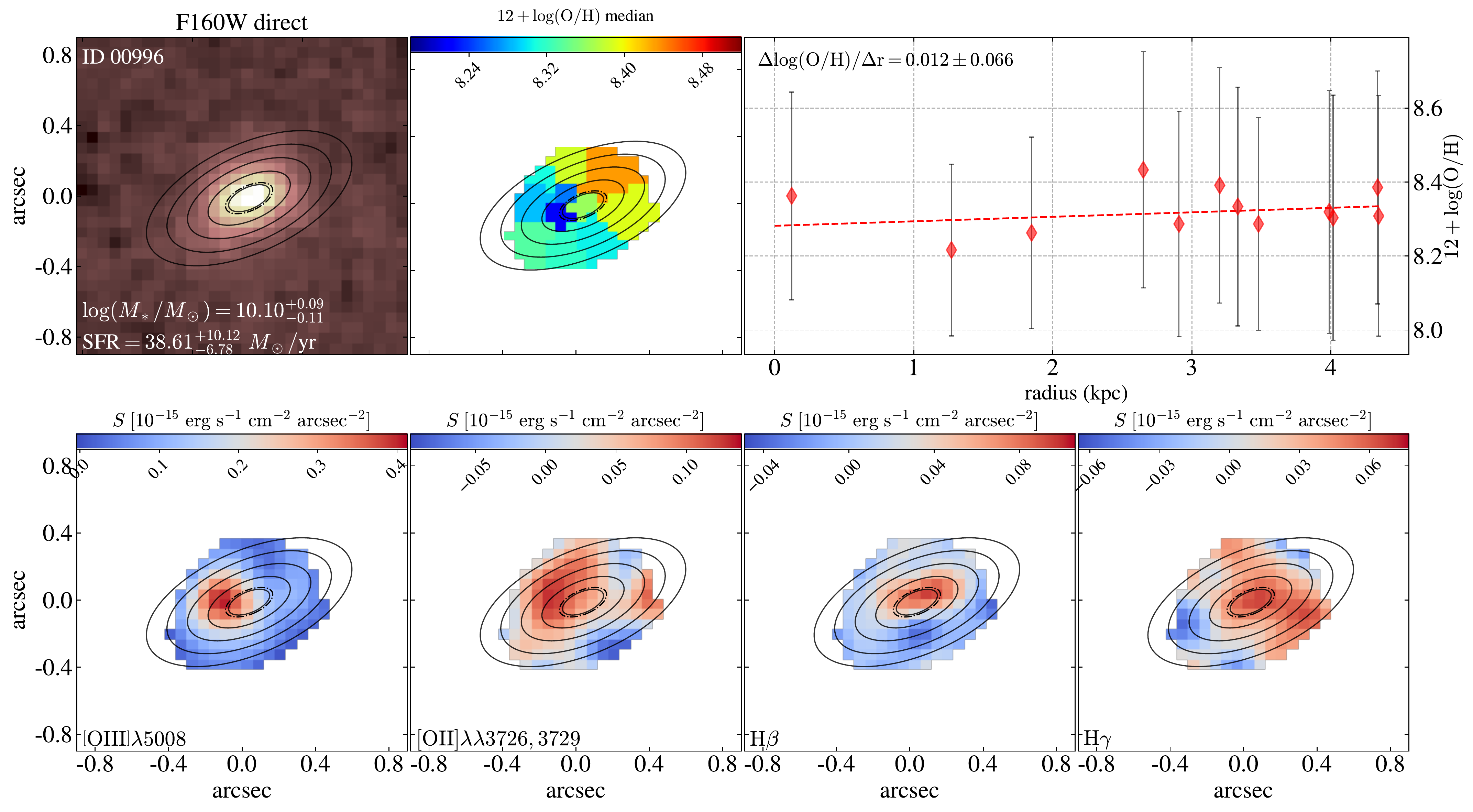}{0.495\textwidth}{ID 00996}
        \fig{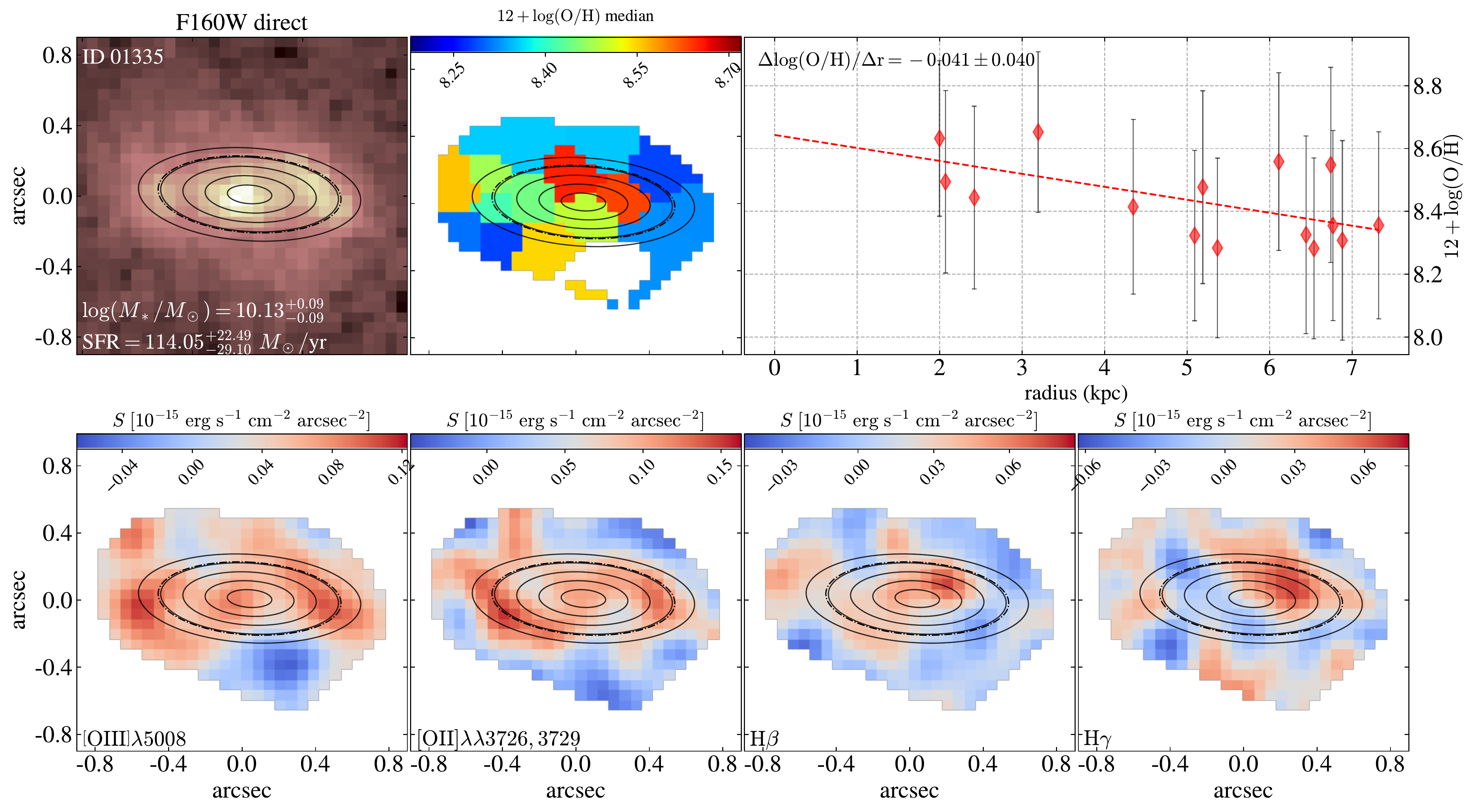}{0.495\textwidth}{ID 01335}
        }

    \gridline{
    \fig{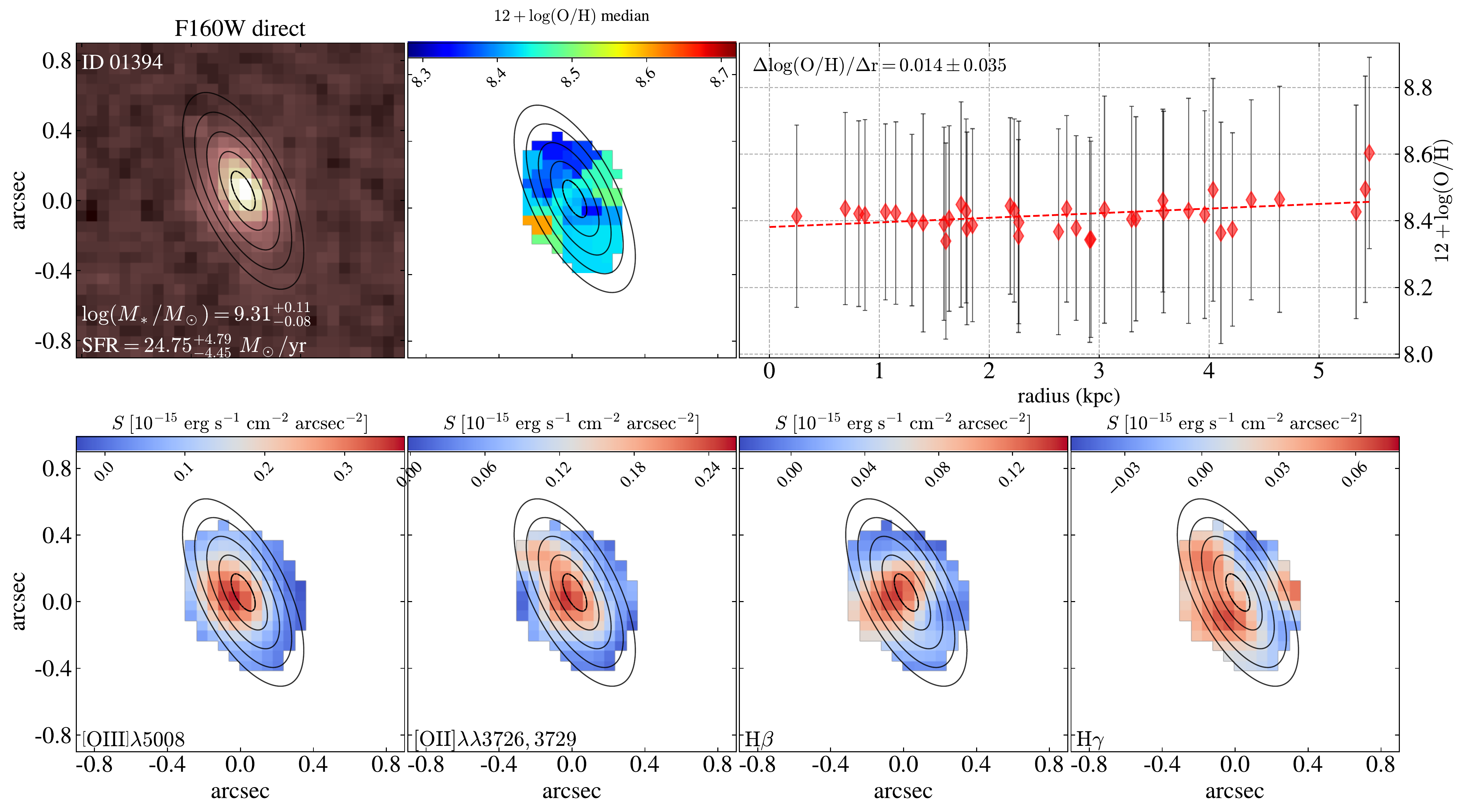}{0.495\textwidth}{ID 01394}
    \fig{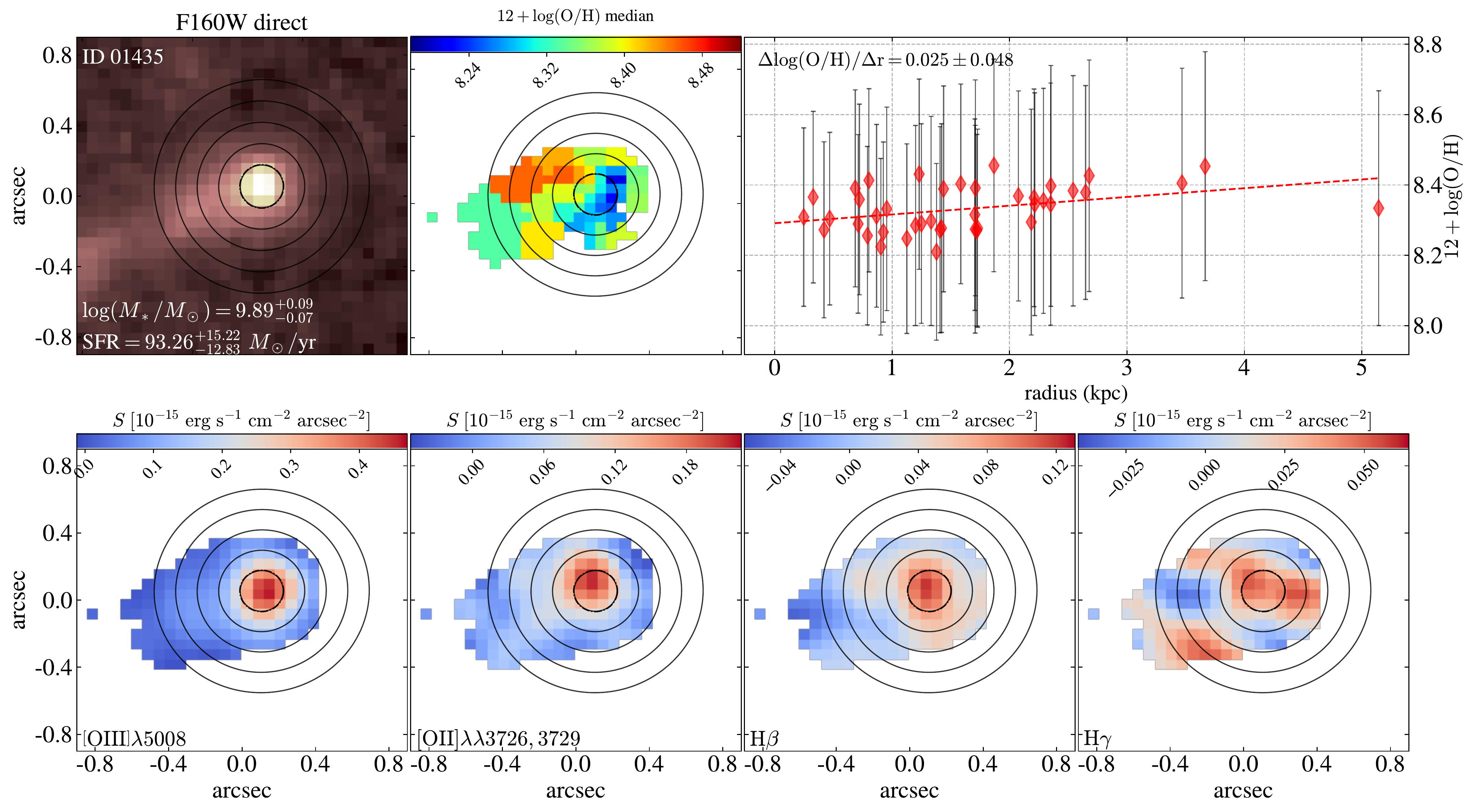}{0.495\textwidth}{ID 01435}
    }
    \gridline{
    \fig{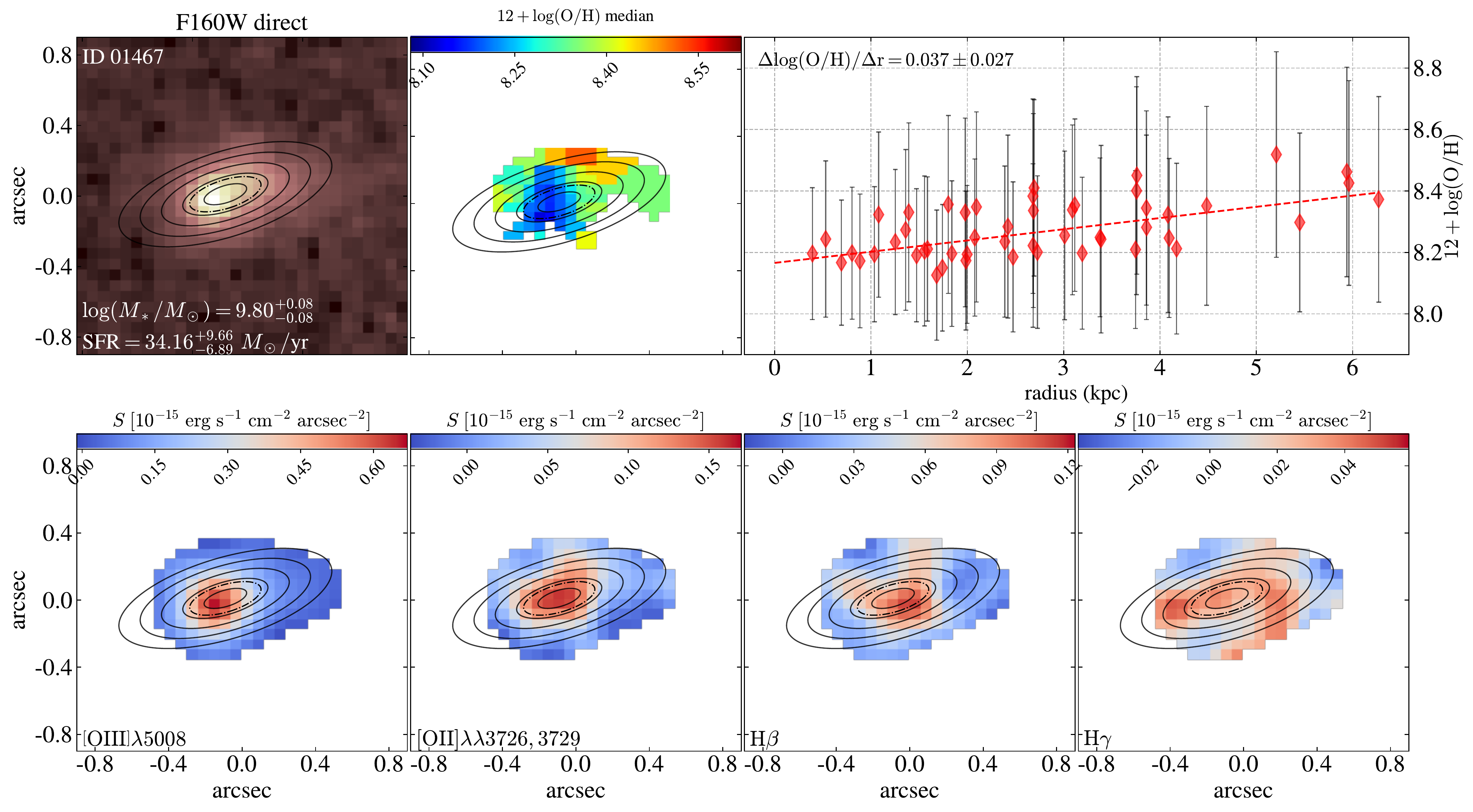}{0.495\textwidth}{ID 01464}
    \fig{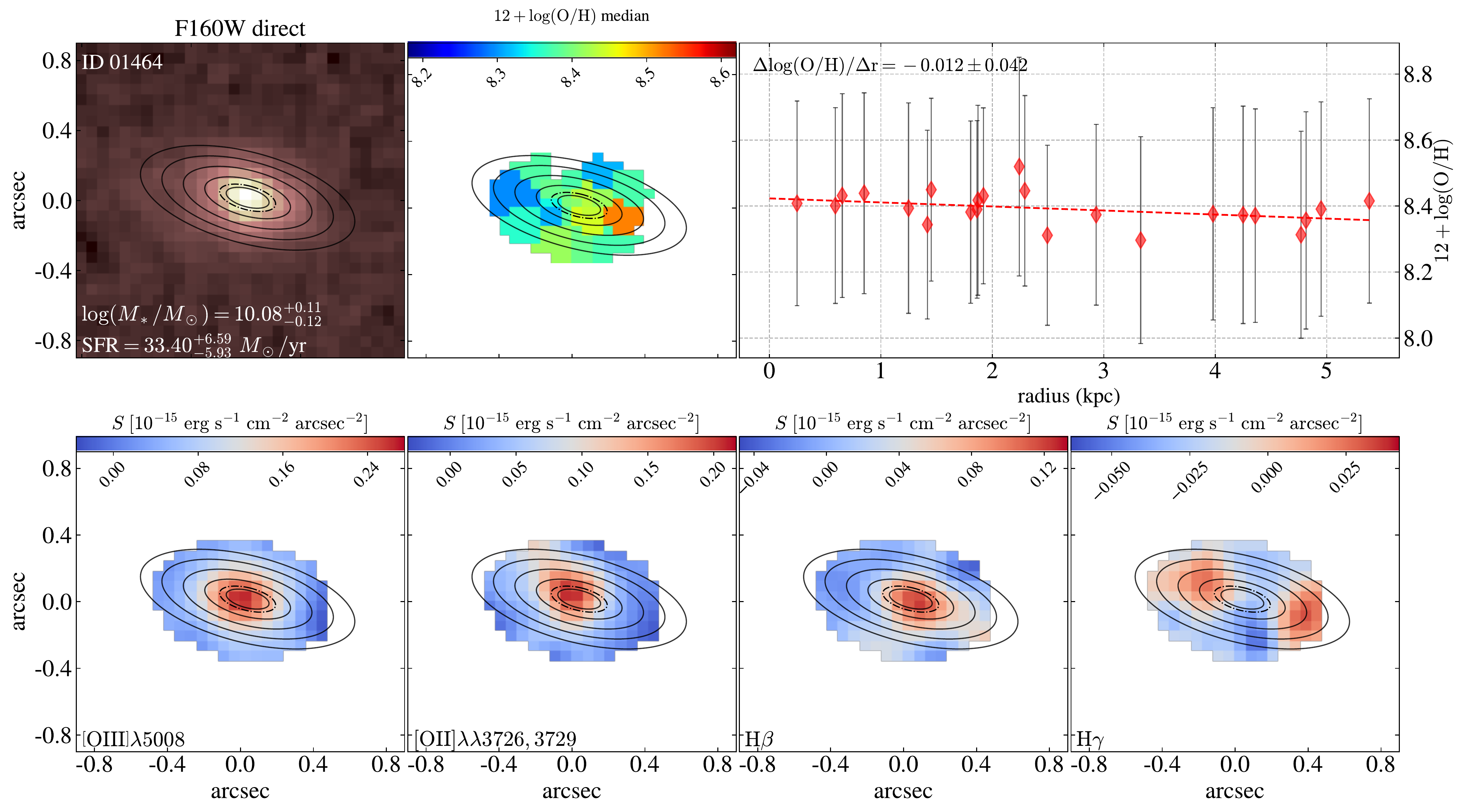}{0.495\textwidth}{ID 01467}
    }
    
\caption{(Continued)}
\end{figure}
\addtocounter{figure}{-1}

\begin{figure}[htbp]
    \centering
    \gridline{
    \fig{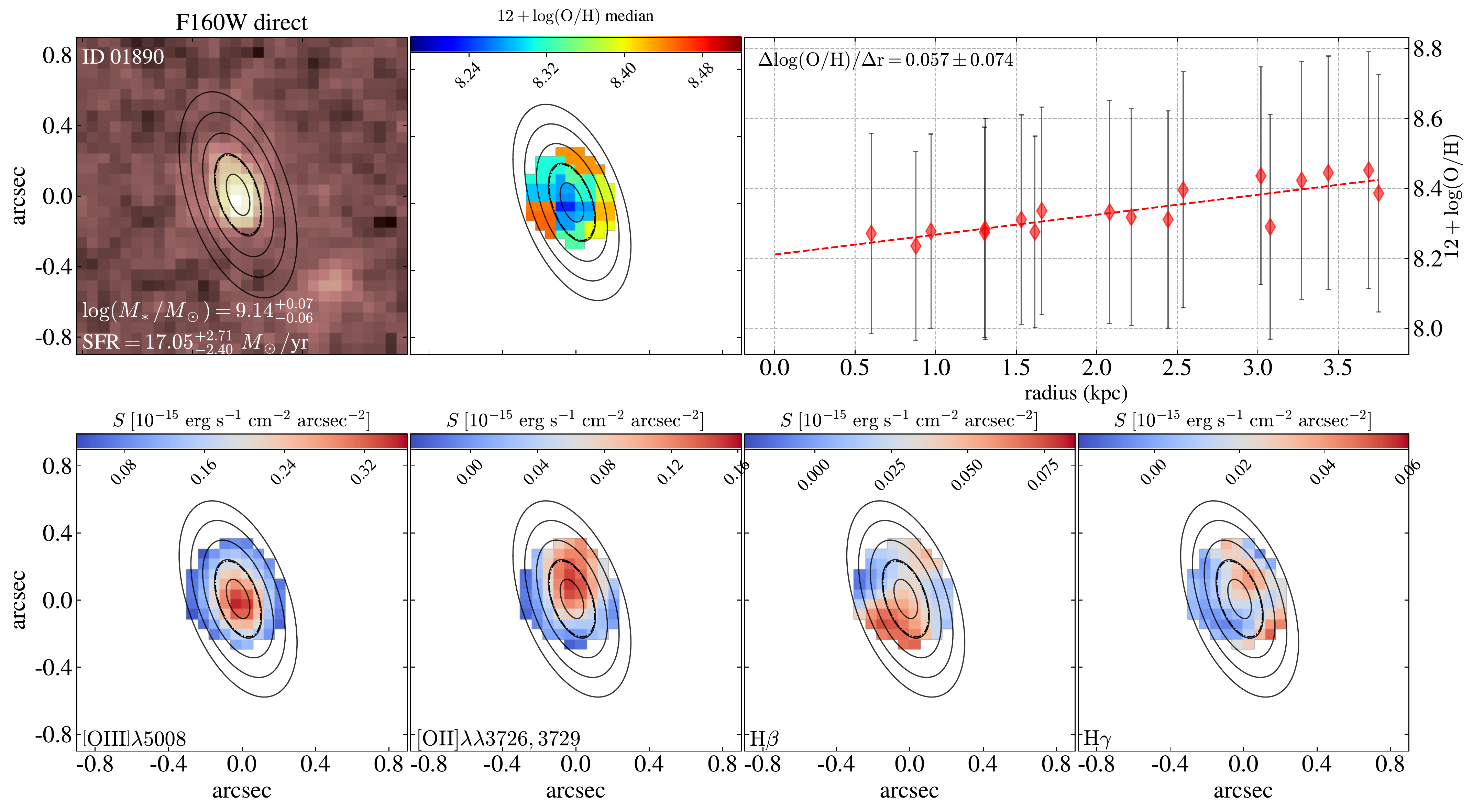}{0.495\textwidth}{ID 01890}
    \fig{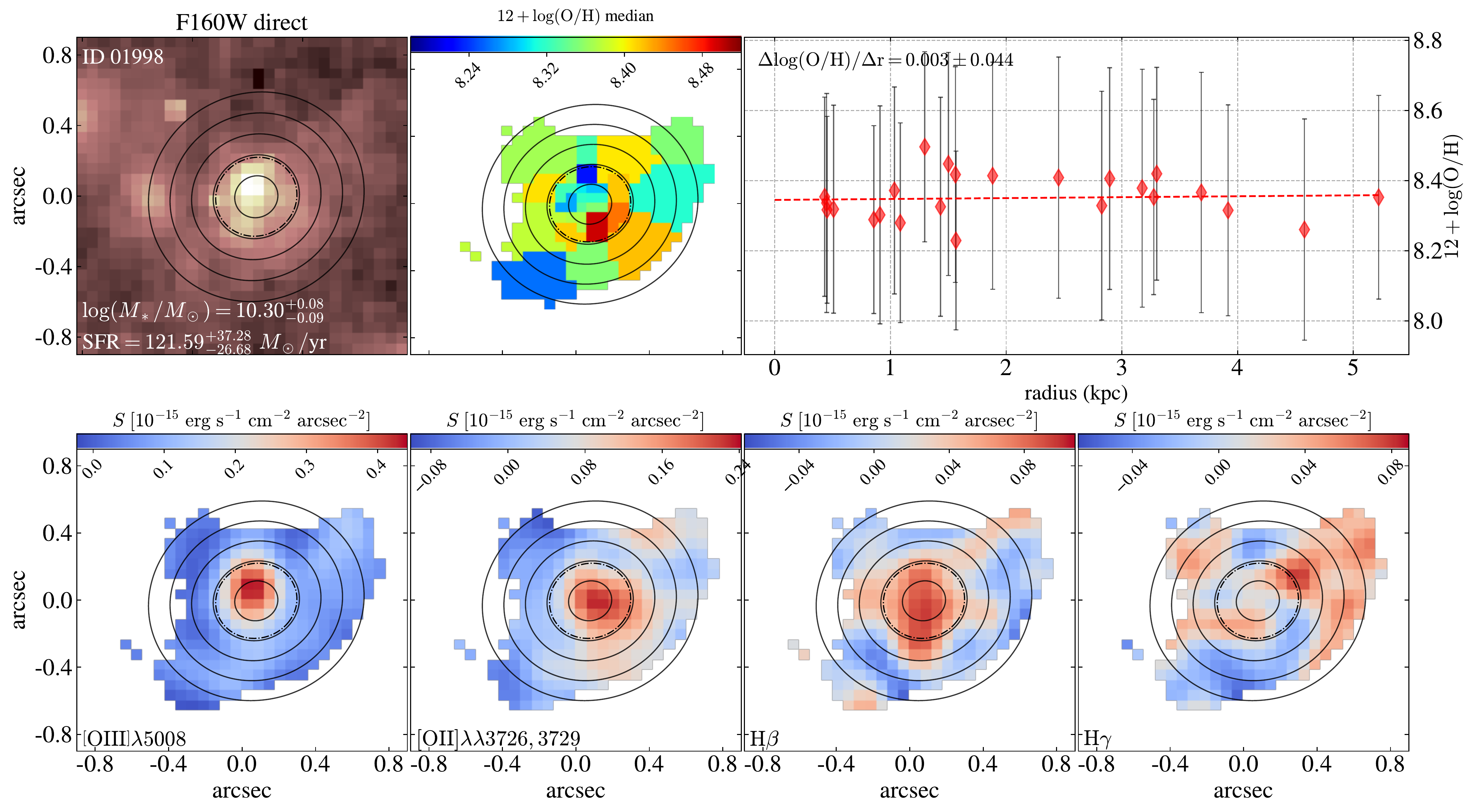}{0.495\textwidth}{ID 01998}}

    \gridline{
    \fig{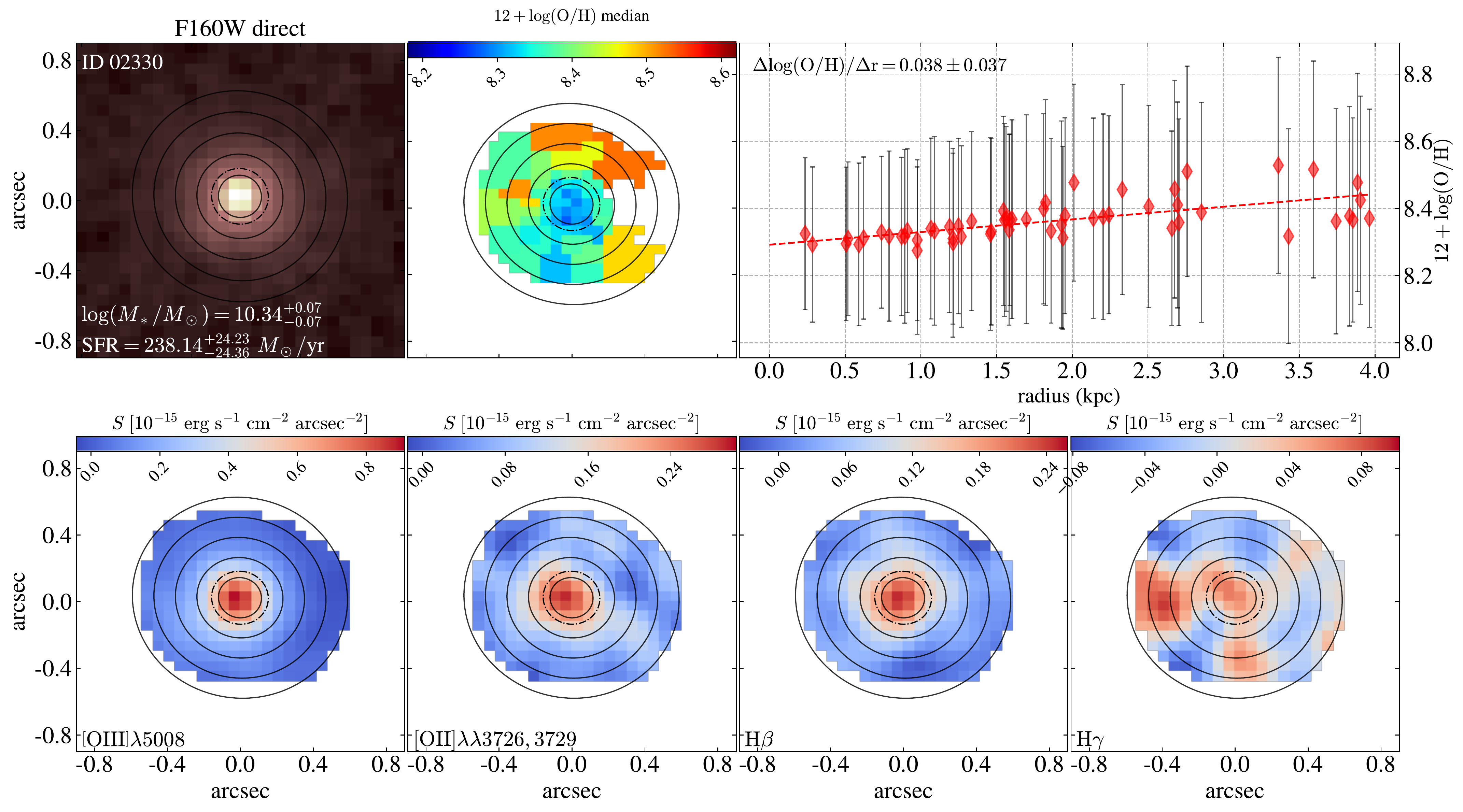}{0.495\textwidth}{ID 02327}
    \fig{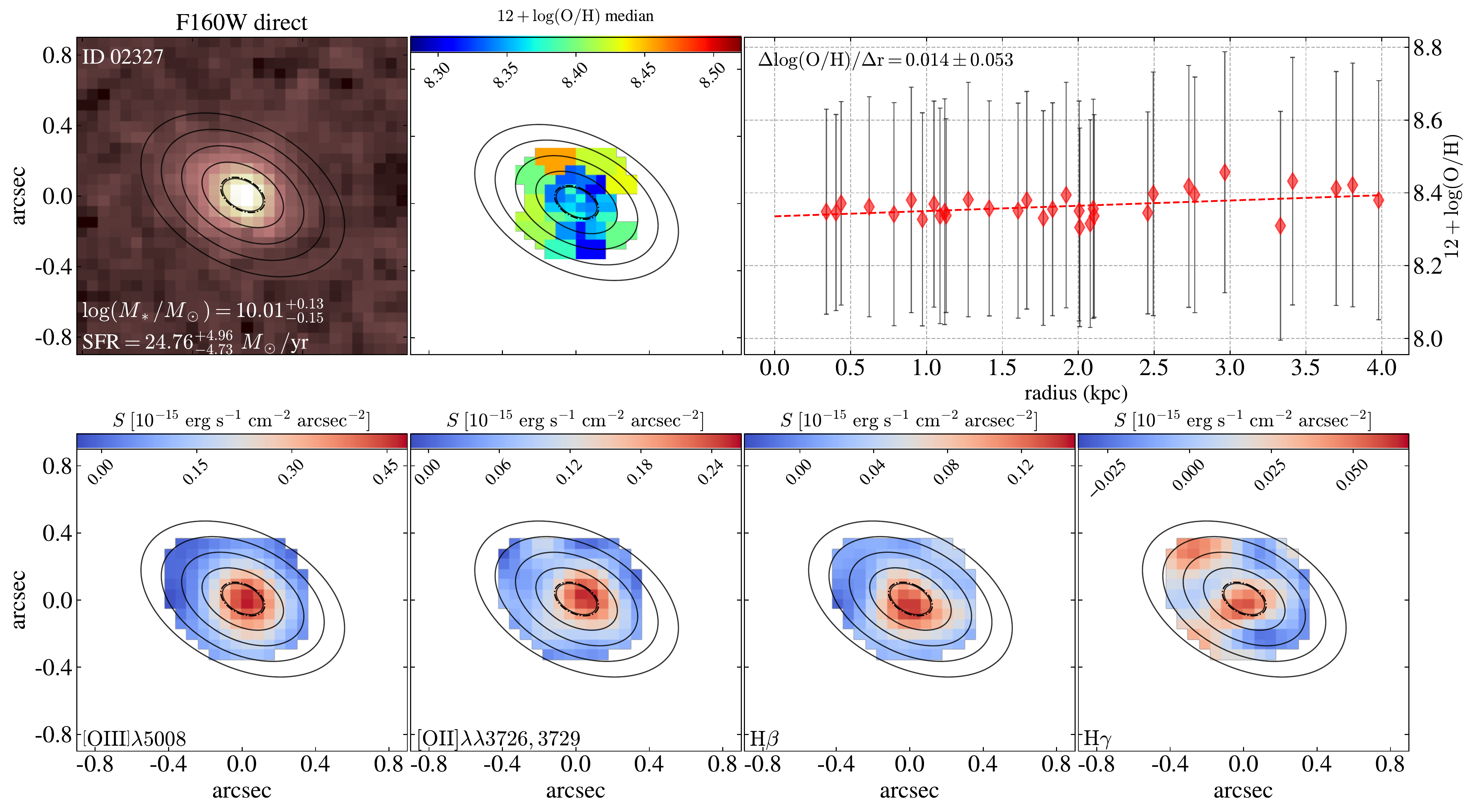}{0.495\textwidth}{ID 02330}}

    \gridline{
    \fig{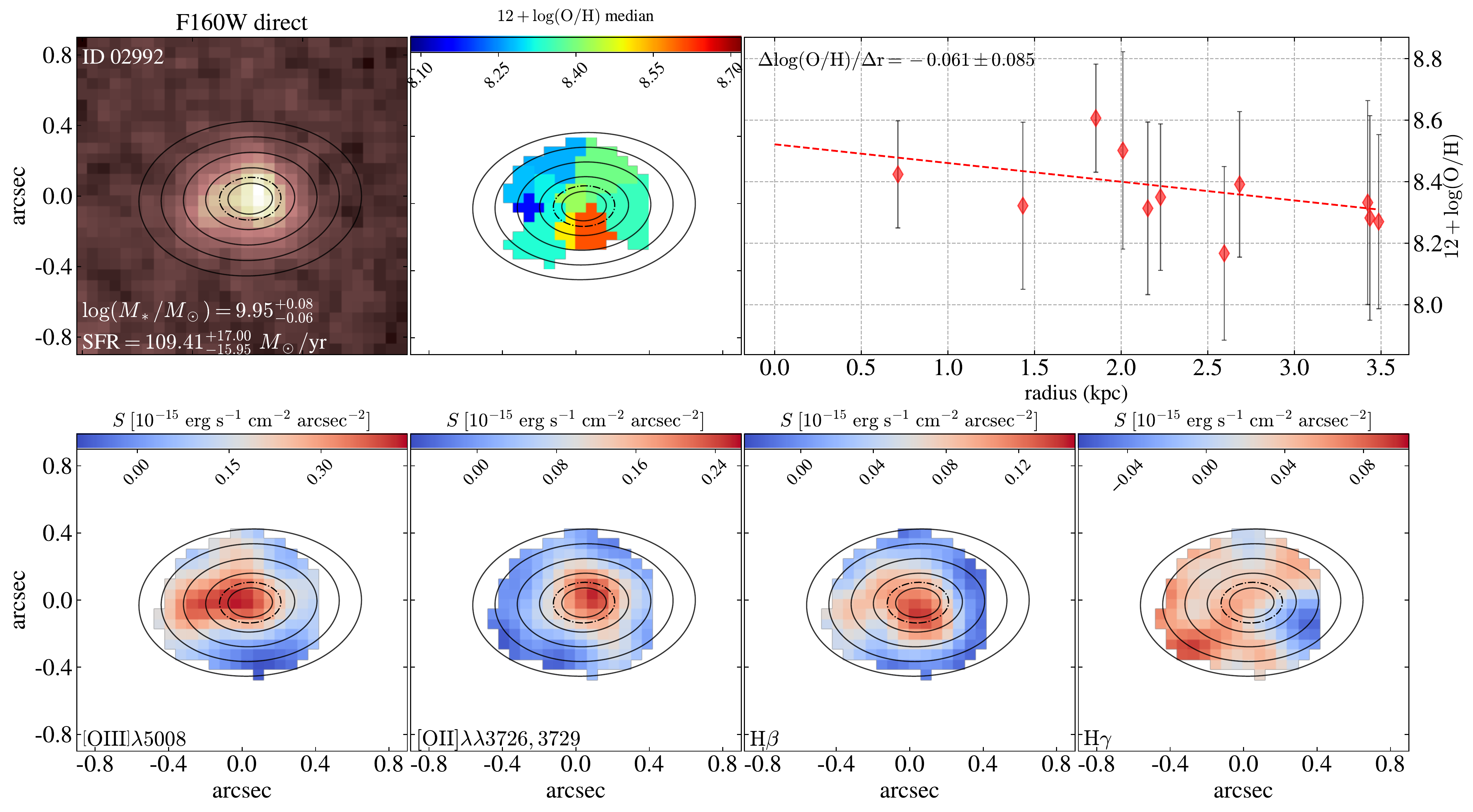}{0.495\textwidth}{ID 02992}
    \fig{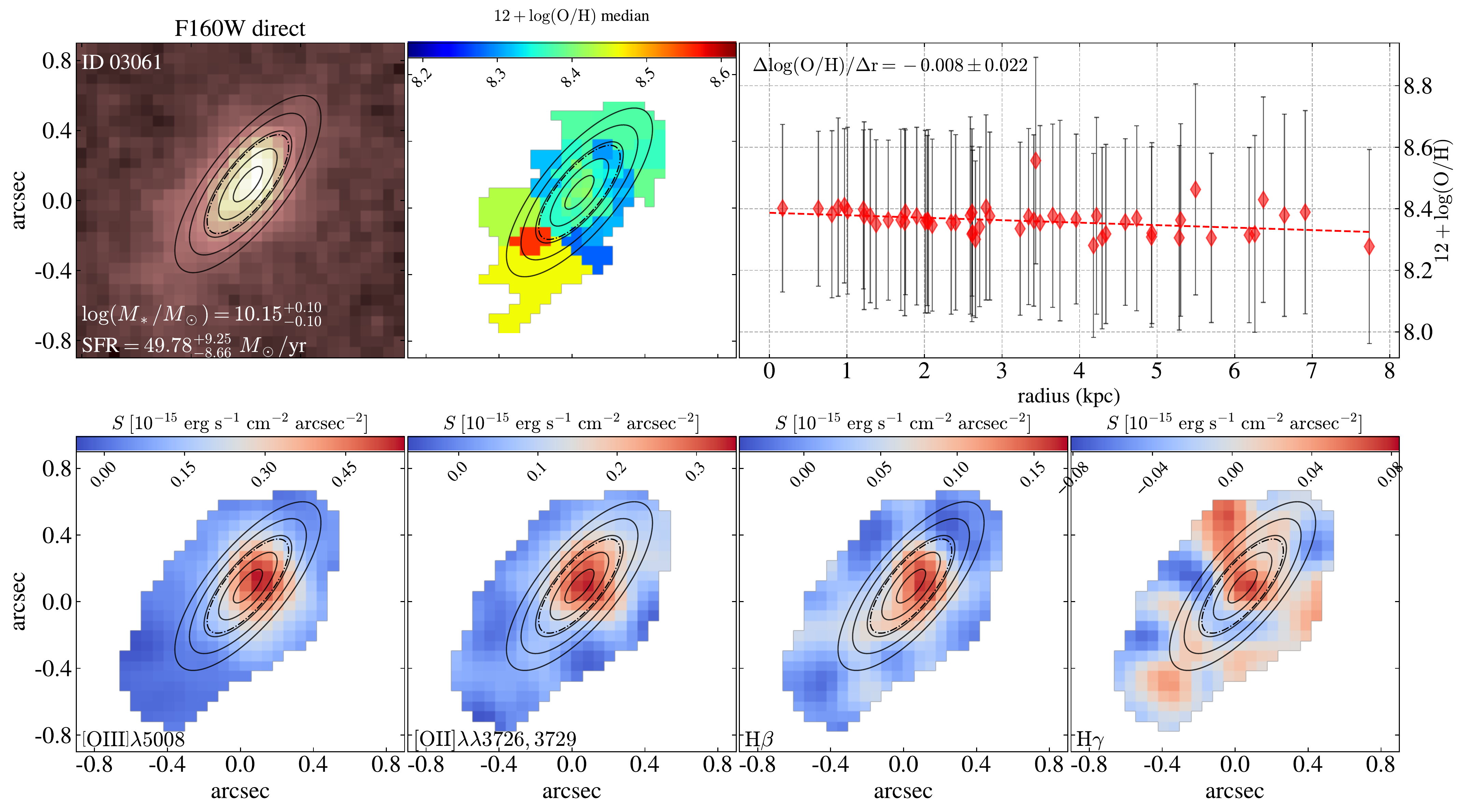}{0.495\textwidth}{ID 03061}}

\caption{(Continued)}
\end{figure}
\addtocounter{figure}{-1}

\begin{figure}[htbp]
    \centering
    
    \gridline{
    \fig{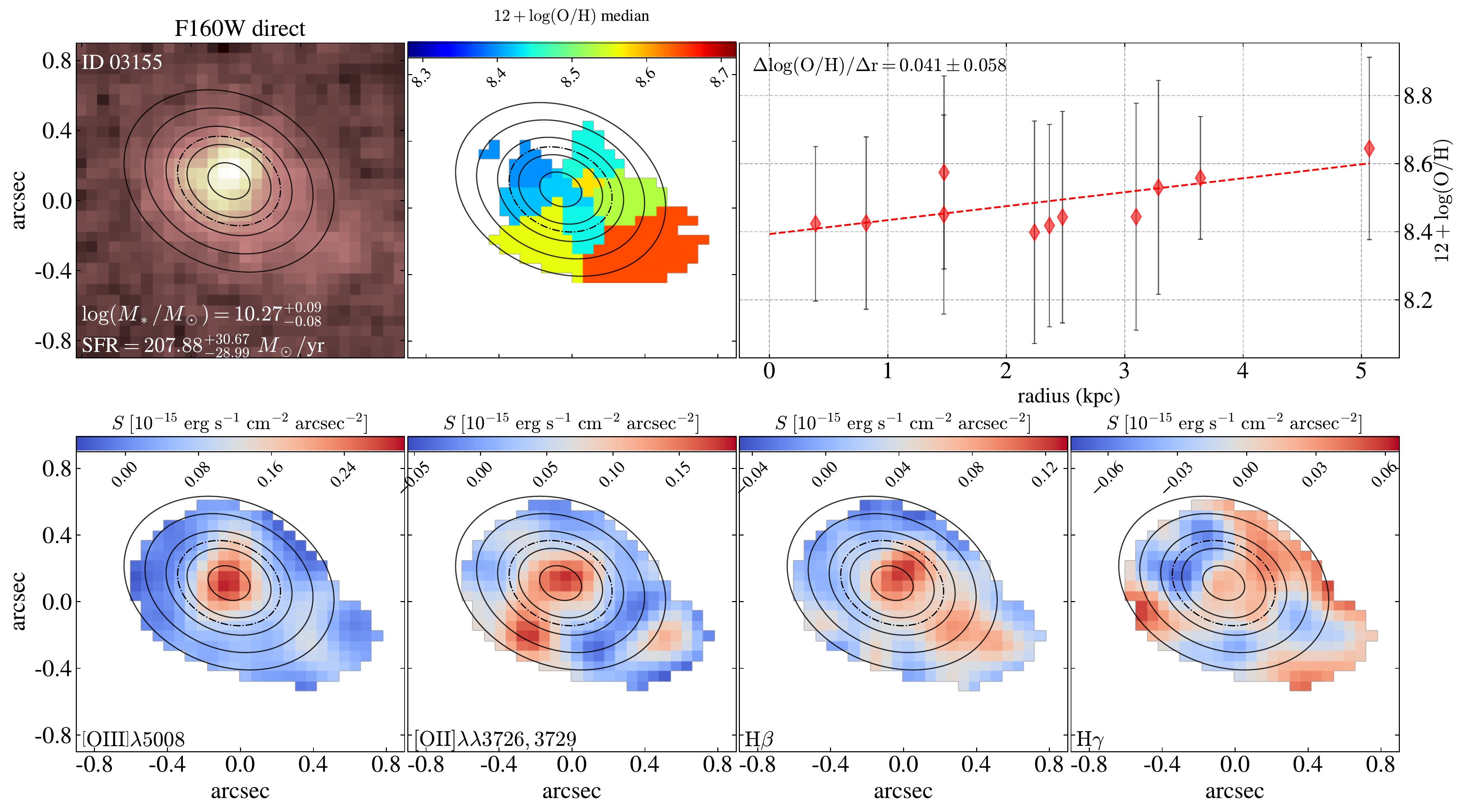}{0.495\textwidth}{ID 03155}
    \fig{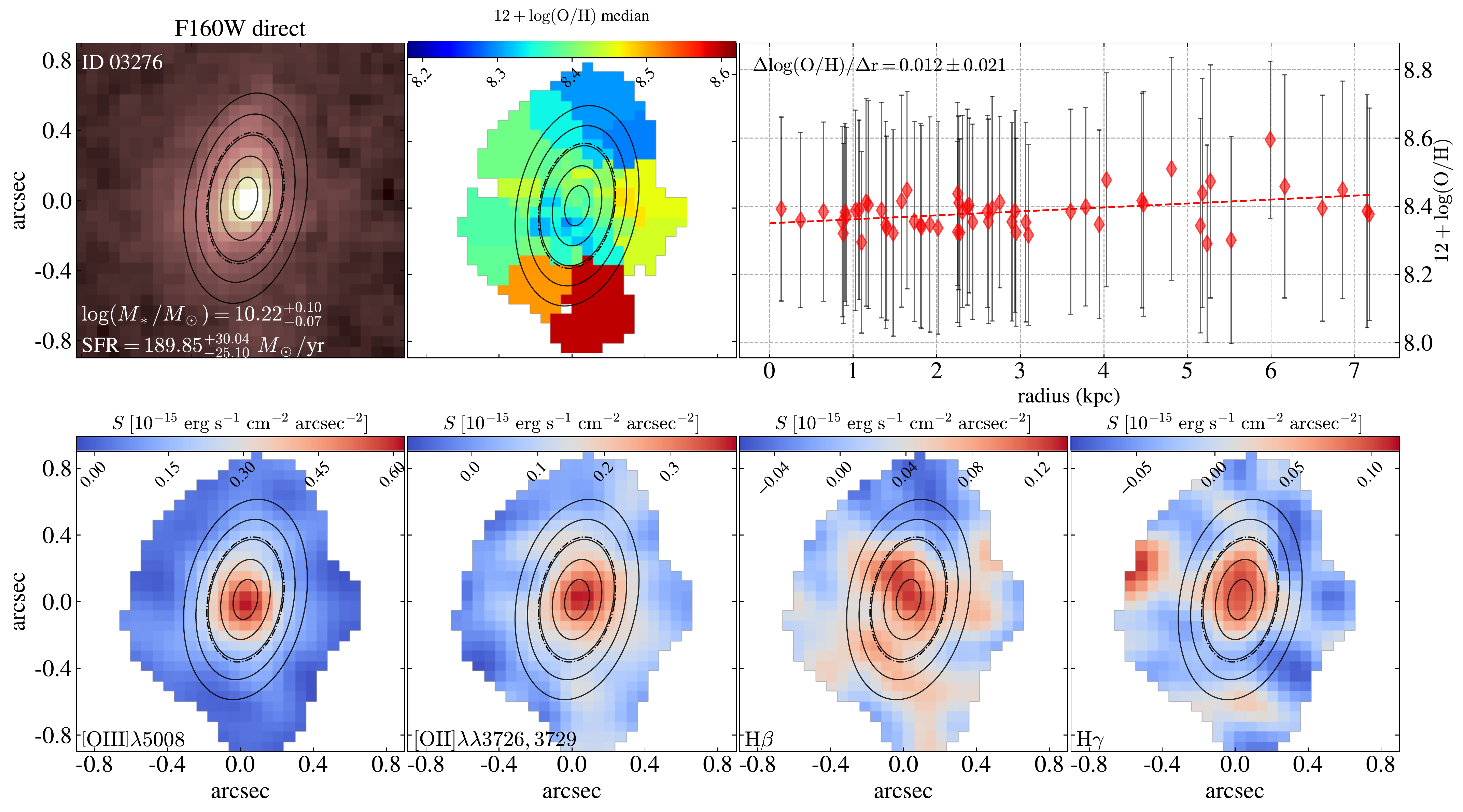}{0.495\textwidth}{ID 03276}}
    \gridline{
    \fig{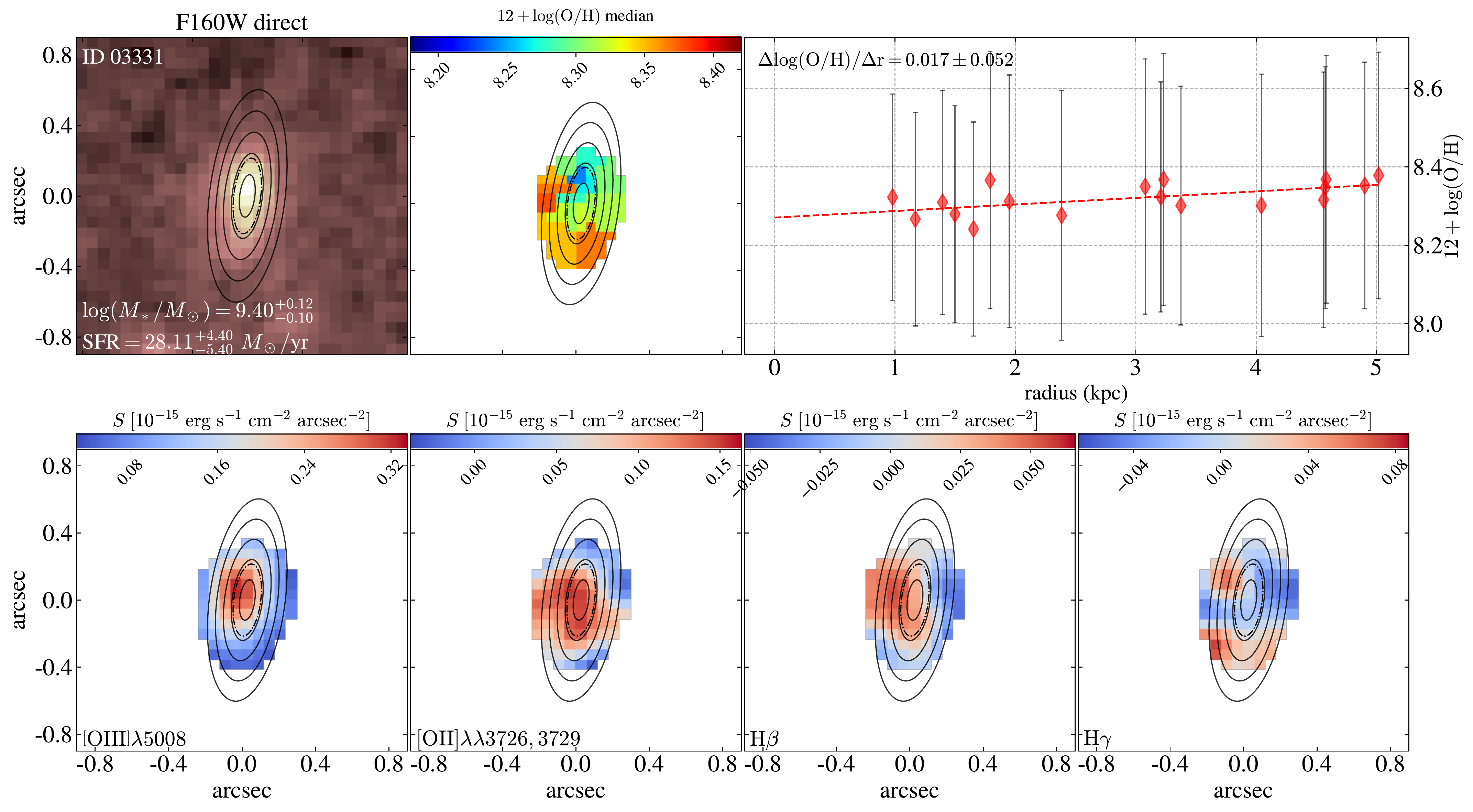}{0.495\textwidth}{ID 03331}
    \fig{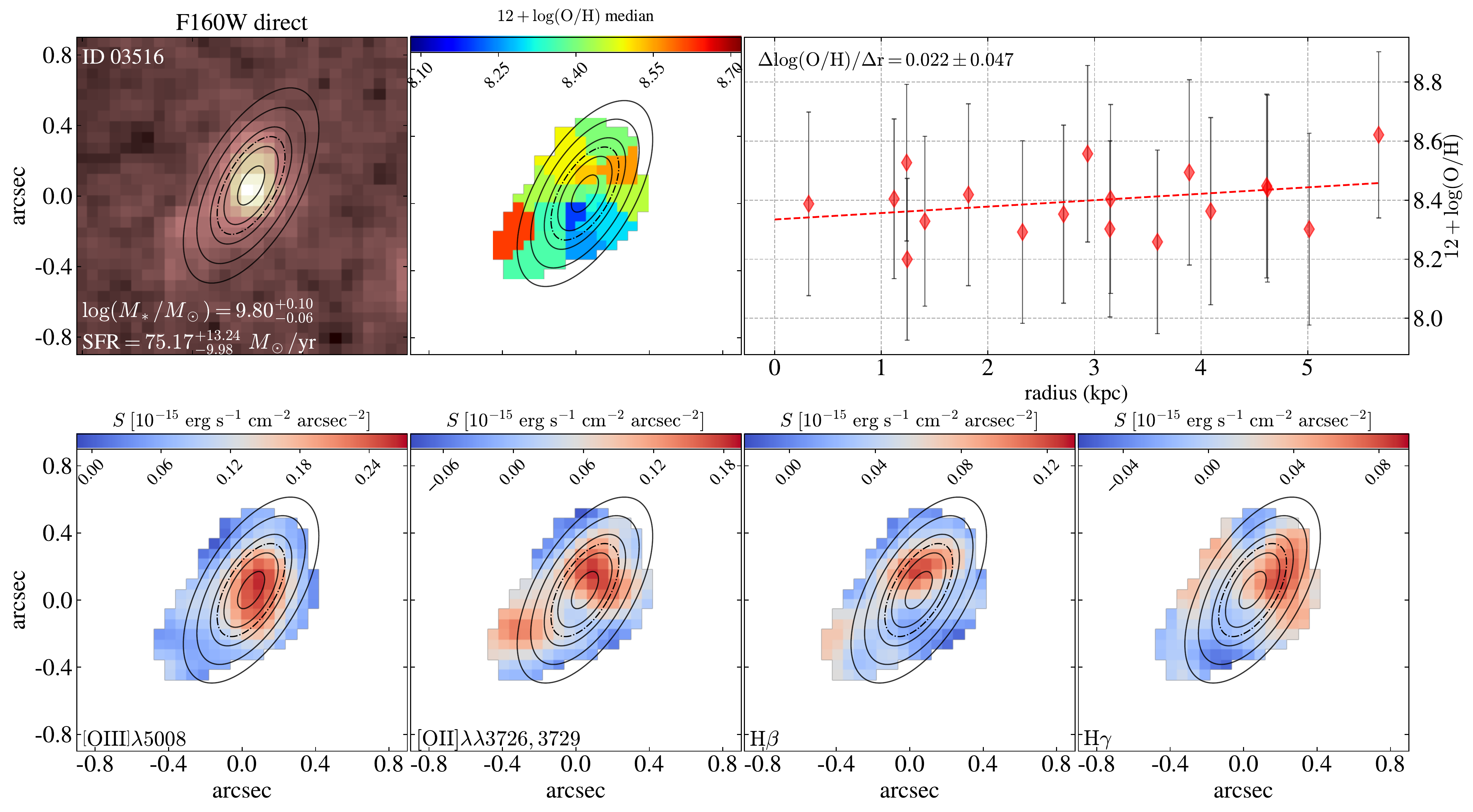}{0.495\textwidth}{ID 03516}}

\caption{(Continued)}
\end{figure}

\bibliography{sample631}{}

\begin{thebibliography}{}
\expandafter\ifx\csname natexlab\endcsname\relax\def\natexlab#1{#1}\fi
\providecommand{\url}[1]{\href{#1}{#1}}
\providecommand{\dodoi}[1]{doi:~\href{http://doi.org/#1}{\nolinkurl{#1}}}
\providecommand{\doeprint}[1]{\href{http://ascl.net/#1}{\nolinkurl{http://ascl.net/#1}}}
\providecommand{\doarXiv}[1]{\href{https://arxiv.org/abs/#1}{\nolinkurl{https://arxiv.org/abs/#1}}}

\bibitem[{{Bian} {et~al.}(2018){Bian}, {Kewley}, \& {Dopita}}]{bian:18}
{Bian}, F., {Kewley}, L.~J., \& {Dopita}, M.~A. 2018, \apj, 859, 175,
  \dodoi{10.3847/1538-4357/aabd74}

\bibitem[{{Brammer} \& {Matharu}(2021)}]{2021zndo...5012699B}
{Brammer}, G., \& {Matharu}, J. 2021, {gbrammer/grizli: Release 2021}, 1.3.2,
  Zenodo, \dodoi{10.5281/zenodo.5012699}

\bibitem[{{Cai} {et~al.}(2016){Cai}, {Fan}, {Peirani}, {Bian}, {Frye},
  {McGreer}, {Prochaska}, {Lau}, {Tejos}, {Ho}, \&
  {Schneider}}]{2016ApJ...833..135C}
{Cai}, Z., {Fan}, X., {Peirani}, S., {et~al.} 2016, \apj, 833, 135,
  \dodoi{10.3847/1538-4357/833/2/135}

\bibitem[{{Cai} {et~al.}(2017){Cai}, {Fan}, {Bian}, {Zabludoff}, {Yang},
  {Prochaska}, {McGreer}, {Zheng}, {Kashikawa}, {Wang}, {Frye}, {Green}, \&
  {Jiang}}]{2017ApJ...839..131C}
{Cai}, Z., {Fan}, X., {Bian}, F., {et~al.} 2017, \apj, 839, 131,
  \dodoi{10.3847/1538-4357/aa6a1a}

\bibitem[{{Calzetti} {et~al.}(2000){Calzetti}, {Armus}, {Bohlin}, {Kinney},
  {Koornneef}, \& {Storchi-Bergmann}}]{2000ApJ...533..682C}
{Calzetti}, D., {Armus}, L., {Bohlin}, R.~C., {et~al.} 2000, \apj, 533, 682,
  \dodoi{10.1086/308692}

\bibitem[{{Cappellari} \& {Copin}(2003)}]{2003MNRAS.342..345C}
{Cappellari}, M., \& {Copin}, Y. 2003, \mnras, 342, 345,
  \dodoi{10.1046/j.1365-8711.2003.06541.x}

\bibitem[{Carton {et~al.}(2018)Carton, Brinchmann, Contini, Epinat, Finley,
  Richard, Patrício, Schaye, Nanayakkara, Weilbacher, \&
  Wisotzki}]{10.1093/mnras/sty1343}
Carton, D., Brinchmann, J., Contini, T., {et~al.} 2018, Monthly Notices of the
  Royal Astronomical Society, 478, 4293, \dodoi{10.1093/mnras/sty1343}

\bibitem[{{Coil} {et~al.}(2015){Coil}, {Aird}, {Reddy}, {Shapley}, {Kriek},
  {Siana}, {Mobasher}, {Freeman}, {Price}, \& {Shivaei}}]{2015ApJ...801...35C}
{Coil}, A.~L., {Aird}, J., {Reddy}, N., {et~al.} 2015, \apj, 801, 35,
  \dodoi{10.1088/0004-637X/801/1/35}

\bibitem[{{Cresci} {et~al.}(2010){Cresci}, {Mannucci}, {Maiolino}, {Marconi},
  {Gnerucci}, \& {Magrini}}]{2010Natur.467..811C}
{Cresci}, G., {Mannucci}, F., {Maiolino}, R., {et~al.} 2010, \nat, 467, 811,
  \dodoi{10.1038/nature09451}

\bibitem[{{Curti} {et~al.}(2020){Curti}, {Maiolino}, {Cirasuolo}, {Mannucci},
  {Williams}, {Auger}, {Mercurio}, {Hayden-Pawson}, {Cresci}, {Marconi},
  {Belfiore}, {Cappellari}, {Cicone}, {Cullen}, {Meneghetti}, {Ota}, {Peng},
  {Pettini}, {Swinbank}, \& {Troncoso}}]{2020MNRAS.492..821C}
{Curti}, M., {Maiolino}, R., {Cirasuolo}, M., {et~al.} 2020, \mnras, 492, 821,
  \dodoi{10.1093/mnras/stz3379}

\bibitem[{{Dekel} \& {Birnboim}(2006)}]{2006MNRAS.368....2D}
{Dekel}, A., \& {Birnboim}, Y. 2006, \mnras, 368, 2,
  \dodoi{10.1111/j.1365-2966.2006.10145.x}

\bibitem[{{Dekel} {et~al.}(2009){Dekel}, {Birnboim}, {Engel}, {Freundlich},
  {Goerdt}, {Mumcuoglu}, {Neistein}, {Pichon}, {Teyssier}, \&
  {Zinger}}]{2009Natur.457..451D}
{Dekel}, A., {Birnboim}, Y., {Engel}, G., {et~al.} 2009, \nat, 457, 451,
  \dodoi{10.1038/nature07648}

\bibitem[{{Finlator} \& {Dav{\'e}}(2008)}]{2008MNRAS.385.2181F}
{Finlator}, K., \& {Dav{\'e}}, R. 2008, \mnras, 385, 2181,
  \dodoi{10.1111/j.1365-2966.2008.12991.x}

\bibitem[{{Foreman-Mackey} {et~al.}(2013){Foreman-Mackey}, {Hogg}, {Lang}, \&
  {Goodman}}]{2013PASP..125..306F}
{Foreman-Mackey}, D., {Hogg}, D.~W., {Lang}, D., \& {Goodman}, J. 2013, \pasp,
  125, 306, \dodoi{10.1086/670067}

\bibitem[{{F{\"o}rster Schreiber} {et~al.}(2018){F{\"o}rster Schreiber},
  {Renzini}, {Mancini}, {Genzel}, {Bouch{\'e}}, {Cresci}, {Hicks}, {Lilly},
  {Peng}, {Burkert}, {Carollo}, {Cimatti}, {Daddi}, {Davies}, {Genel}, {Kurk},
  {Lang}, {Lutz}, {Mainieri}, {McCracken}, {Mignoli}, {Naab}, {Oesch},
  {Pozzetti}, {Scodeggio}, {Shapiro Griffin}, {Shapley}, {Sternberg},
  {Tacchella}, {Tacconi}, {Wuyts}, \& {Zamorani}}]{Forster2018}
{F{\"o}rster Schreiber}, N.~M., {Renzini}, A., {Mancini}, C., {et~al.} 2018,
  \apjs, 238, 21, \dodoi{10.3847/1538-4365/aadd49}

\bibitem[{{Hemler} {et~al.}(2021){Hemler}, {Torrey}, {Qi}, {Hernquist},
  {Vogelsberger}, {Ma}, {Kewley}, {Nelson}, {Pillepich}, {Pakmor}, \&
  {Marinacci}}]{2021MNRAS.506.3024H}
{Hemler}, Z.~S., {Torrey}, P., {Qi}, J., {et~al.} 2021, \mnras, 506, 3024,
  \dodoi{10.1093/mnras/stab1803}

\bibitem[{Henry {et~al.}(2021)Henry, Rafelski, Sunnquist, Pirzkal, Pacifici,
  Atek, Bagley, Baronchelli, Barro, Bunker, Colbert, Dai, Elmegreen, Elmegreen,
  Finkelstein, Kocevski, Koekemoer, Malkan, Martin, Mehta, Pahl, Papovich,
  Rutkowski, Almeida, Scarlata, Snyder, \& Teplitz}]{Henry:2021ju}
Henry, A.~L., Rafelski, M.~A., Sunnquist, B., {et~al.} 2021, ApJ, 919, 143,
  \dodoi{10.3847/1538-4357/ac1105}

\bibitem[{{Ho} {et~al.}(2015){Ho}, {Kudritzki}, {Kewley}, {Zahid}, {Dopita},
  {Bresolin}, \& {Rupke}}]{2015MNRAS.448.2030H}
{Ho}, I.~T., {Kudritzki}, R.-P., {Kewley}, L.~J., {et~al.} 2015, \mnras, 448,
  2030, \dodoi{10.1093/mnras/stv067}

\bibitem[{{Ho} {et~al.}(2017){Ho}, {Seibert}, {Meidt}, {Kudritzki},
  {Kobayashi}, {Groves}, {Kewley}, {Madore}, {Rich}, {Schinnerer},
  {D'Agostino}, \& {Poetrodjojo}}]{2017ApJ...846...39H}
{Ho}, I.~T., {Seibert}, M., {Meidt}, S.~E., {et~al.} 2017, \apj, 846, 39,
  \dodoi{10.3847/1538-4357/aa8460}

\bibitem[{{Hou} {et~al.}(2000){Hou}, {Prantzos}, \&
  {Boissier}}]{2000A&A...362..921H}
{Hou}, J.~L., {Prantzos}, N., \& {Boissier}, S. 2000, \aap, 362, 921.
\newblock \doarXiv{astro-ph/0007164}

\bibitem[{Jones {et~al.}(2013)Jones, Ellis, Richard, \& Jullo}]{jones2013}
Jones, T., Ellis, R.~S., Richard, J., \& Jullo, E. 2013, \apj, 765, 48,
  \dodoi{10.1088/0004-637x/765/1/48}

\bibitem[{Jones {et~al.}(2010)Jones, Swinbank, Ellis, Richard, \&
  Stark}]{2010MNRAS.404.1247J}
Jones, T.~A., Swinbank, A.~M., Ellis, R.~S., Richard, J., \& Stark, D.~P. 2010,
  MNRAS, 404, 1247

\bibitem[{{Juneau} {et~al.}(2014){Juneau}, {Bournaud}, {Charlot}, {Daddi},
  {Elbaz}, {Trump}, {Brinchmann}, {Dickinson}, {Duc}, {Gobat}, {Jean-Baptiste},
  {Le Floc'h}, {Lehnert}, {Pacifici}, {Pannella}, \&
  {Schreiber}}]{2014ApJ...788...88J}
{Juneau}, S., {Bournaud}, F., {Charlot}, S., {et~al.} 2014, \apj, 788, 88,
  \dodoi{10.1088/0004-637X/788/1/88}

\bibitem[{{Kelly}(2007)}]{2007ApJ...665.1489K}
{Kelly}, B.~C. 2007, \apj, 665, 1489, \dodoi{10.1086/519947}

\bibitem[{{Kere{\v{s}}} {et~al.}(2009){Kere{\v{s}}}, {Katz}, {Fardal},
  {Dav{\'e}}, \& {Weinberg}}]{Keres2009}
{Kere{\v{s}}}, D., {Katz}, N., {Fardal}, M., {Dav{\'e}}, R., \& {Weinberg},
  D.~H. 2009, \mnras, 395, 160, \dodoi{10.1111/j.1365-2966.2009.14541.x}

\bibitem[{{Kere{\v{s}}} {et~al.}(2005){Kere{\v{s}}}, {Katz}, {Weinberg}, \&
  {Dav{\'e}}}]{2005MNRAS.363....2K}
{Kere{\v{s}}}, D., {Katz}, N., {Weinberg}, D.~H., \& {Dav{\'e}}, R. 2005,
  \mnras, 363, 2, \dodoi{10.1111/j.1365-2966.2005.09451.x}

\bibitem[{{Kleiner} {et~al.}(2017){Kleiner}, {Pimbblet}, {Jones}, {Koribalski},
  \& {Serra}}]{2017MNRAS.466.4692K}
{Kleiner}, D., {Pimbblet}, K.~A., {Jones}, D.~H., {Koribalski}, B.~S., \&
  {Serra}, P. 2017, \mnras, 466, 4692, \dodoi{10.1093/mnras/stw3328}

\bibitem[{{Leethochawalit} {et~al.}(2016){Leethochawalit}, {Jones}, {Ellis},
  {Stark}, {Richard}, {Zitrin}, \& {Auger}}]{Leethochawalit2016}
{Leethochawalit}, N., {Jones}, T.~A., {Ellis}, R.~S., {et~al.} 2016, \apj, 820,
  84, \dodoi{10.3847/0004-637X/820/2/84}

\bibitem[{{Ma} {et~al.}(2017){Ma}, {Hopkins}, {Feldmann}, {Torrey},
  {Faucher-Gigu{\`e}re}, \& {Kere{\v{s}}}}]{2017MNRAS.466.4780M}
{Ma}, X., {Hopkins}, P.~F., {Feldmann}, R., {et~al.} 2017, \mnras, 466, 4780,
  \dodoi{10.1093/mnras/stx034}

\bibitem[{Muratov {et~al.}(2015)Muratov, Kereš, Faucher-Giguère, Hopkins,
  Quataert, \& Murray}]{Muratov:2015er}
Muratov, A.~L., Kereš, D., Faucher-Giguère, C.-A., {et~al.} 2015, MNRAS, 454,
  2691 , \dodoi{10.1093/mnras/stv2126}

\bibitem[{{Pagel} \& {Edmunds}(1981)}]{1981ARA&A..19...77P}
{Pagel}, B.~E.~J., \& {Edmunds}, M.~G. 1981, \araa, 19, 77,
  \dodoi{10.1146/annurev.aa.19.090181.000453}

\bibitem[{{Peng} {et~al.}(2002){Peng}, {Ho}, {Impey}, \&
  {Rix}}]{2002AJ....124..266P}
{Peng}, C.~Y., {Ho}, L.~C., {Impey}, C.~D., \& {Rix}, H.-W. 2002, \aj, 124,
  266, \dodoi{10.1086/340952}

\bibitem[{Peng \& Maiolino(2014)}]{10.1093/mnras/stu1288}
Peng, Y.-j., \& Maiolino, R. 2014, Monthly Notices of the Royal Astronomical
  Society, 443, 3643, \dodoi{10.1093/mnras/stu1288}

\bibitem[{{Rupke} {et~al.}(2010){Rupke}, {Kewley}, \&
  {Barnes}}]{2010ApJ...710L.156R}
{Rupke}, D. S.~N., {Kewley}, L.~J., \& {Barnes}, J.~E. 2010, \apjl, 710, L156,
  \dodoi{10.1088/2041-8205/710/2/L156}

\bibitem[{Salim {et~al.}(2015)Salim, Lee, Dav{\'{e}}, \&
  Dickinson}]{Salim_2015}
Salim, S., Lee, J.~C., Dav{\'{e}}, R., \& Dickinson, M. 2015, The Astrophysical
  Journal, 808, 25, \dodoi{10.1088/0004-637x/808/1/25}

\bibitem[{{Sanders} {et~al.}(2021){Sanders}, {Shapley}, {Jones}, {Reddy},
  {Kriek}, {Siana}, {Coil}, {Mobasher}, {Shivaei}, {Dav{\'e}}, {Azadi},
  {Price}, {Leung}, {Freeman}, {Fetherolf}, {de Groot}, {Zick}, \&
  {Barro}}]{2021ApJ...914...19S}
{Sanders}, R.~L., {Shapley}, A.~E., {Jones}, T., {et~al.} 2021, \apj, 914, 19,
  \dodoi{10.3847/1538-4357/abf4c1}

\bibitem[{{Searle}(1971)}]{1971ApJ...168..327S}
{Searle}, L. 1971, \apj, 168, 327, \dodoi{10.1086/151090}

\bibitem[{{Shi} {et~al.}(2021){Shi}, {Cai}, {Fan}, {Zheng}, {Huang}, \&
  {Xu}}]{Shi2021}
{Shi}, D.~D., {Cai}, Z., {Fan}, X., {et~al.} 2021, \apj, 915, 32,
  \dodoi{10.3847/1538-4357/abfec0}

\bibitem[{Simons {et~al.}(2021)Simons, Papovich, Momcheva, Trump, Brammer,
  Estrada-Carpenter, Backhaus, Cleri, Finkelstein, Giavalisco, Ji, Jung,
  Matharu, \& Weiner}]{Simons_2021}
Simons, R.~C., Papovich, C., Momcheva, I., {et~al.} 2021, The Astrophysical
  Journal, 923, 203, \dodoi{10.3847/1538-4357/ac28f4}

\bibitem[{{Stott} {et~al.}(2014){Stott}, {Sobral}, {Swinbank}, {Smail},
  {Bower}, {Best}, {Sharples}, {Geach}, \& {Matthee}}]{2014MNRAS.443.2695S}
{Stott}, J.~P., {Sobral}, D., {Swinbank}, A.~M., {et~al.} 2014, \mnras, 443,
  2695, \dodoi{10.1093/mnras/stu1343}

\bibitem[{{Swinbank} {et~al.}(2012){Swinbank}, {Sobral}, {Smail}, {Geach},
  {Best}, {McCarthy}, {Crain}, \& {Theuns}}]{Swinbank2012}
{Swinbank}, A.~M., {Sobral}, D., {Smail}, I., {et~al.} 2012, \mnras, 426, 935,
  \dodoi{10.1111/j.1365-2966.2012.21774.x}

\bibitem[{{Tissera} {et~al.}(2019){Tissera}, {Rosas-Guevara}, {Bower}, {Crain},
  {del P Lagos}, {Schaller}, {Schaye}, \& {Theuns}}]{2019MNRAS.482.2208T}
{Tissera}, P.~B., {Rosas-Guevara}, Y., {Bower}, R.~G., {et~al.} 2019, \mnras,
  482, 2208, \dodoi{10.1093/mnras/sty2817}

\bibitem[{{Wang} {et~al.}(2017){Wang}, {Jones}, {Treu}, {Morishita},
  {Abramson}, {Brammer}, {Huang}, {Malkan}, {Schmidt}, {Fontana}, {Grillo},
  {Henry}, {Karman}, {Kelly}, {Mason}, {Mercurio}, {Rosati}, {Sharon},
  {Trenti}, \& {Vulcani}}]{2017ApJ...837...89W}
{Wang}, X., {Jones}, T.~A., {Treu}, T., {et~al.} 2017, \apj, 837, 89,
  \dodoi{10.3847/1538-4357/aa603c}

\bibitem[{{Wang} {et~al.}(2019){Wang}, {Jones}, {Treu}, {Hirtenstein},
  {Brammer}, {Daddi}, {Meng}, {Morishita}, {Abramson}, {Henry}, {Peng},
  {Schmidt}, {Sharon}, {Trenti}, \& {Vulcani}}]{2019ApJ...882...94W}
---. 2019, \apj, 882, 94, \dodoi{10.3847/1538-4357/ab3861}

\bibitem[{{Wang} {et~al.}(2020){Wang}, {Jones}, {Treu}, {Daddi}, {Brammer},
  {Sharon}, {Morishita}, {Abramson}, {Colbert}, {Henry}, {Hopkins}, {Malkan},
  {Schmidt}, {Teplitz}, \& {Vulcani}}]{Wang20}
---. 2020, \apj, 900, 183, \dodoi{10.3847/1538-4357/abacce}

\bibitem[{{Wang} {et~al.}(2022){Wang}, {Li}, {Cai}, {Shi}, {Fan}, {Zheng},
  {Bian}, {Teplitz}, {Alavi}, {Colbert}, {Henry}, \&
  {Malkan}}]{2022ApJ...926...70W}
{Wang}, X., {Li}, Z., {Cai}, Z., {et~al.} 2022, \apj, 926, 70,
  \dodoi{10.3847/1538-4357/ac3974}

\bibitem[{Wuyts {et~al.}(2016)Wuyts, Wisnioski, Fossati, Schreiber, Genzel,
  Davies, Mendel, Naab, Röttgers, Wilman, Wuyts, Bandara, Beifiori, Belli,
  Bender, Brammer, Burkert, Chan, Galametz, Kulkarni, Lang, Lutz, Momcheva,
  Nelson, Rosario, Saglia, Seitz, Tacconi, ichi Tadaki, Übler, \& van
  Dokkum}]{Wuyts_2016}
Wuyts, E., Wisnioski, E., Fossati, M., {et~al.} 2016, The Astrophysical
  Journal, 827, 74, \dodoi{10.3847/0004-637x/827/1/74}

\bibitem[{{Zanella} {et~al.}(2015){Zanella}, {Daddi}, {Le Floc'h}, {Bournaud},
  {Gobat}, {Valentino}, {Strazzullo}, {Cibinel}, {Onodera}, {Perret}, {Renaud},
  \& {Vignali}}]{2015Natur.521...54Z}
{Zanella}, A., {Daddi}, E., {Le Floc'h}, E., {et~al.} 2015, \nat, 521, 54,
  \dodoi{10.1038/nature14409}

\end{thebibliography}
\bibliographystyle{aasjournal}

\end{document}